\documentclass[11pt,a4paper]{article}
\input amssymb.sty
\usepackage[dvips]{graphics}
\begin{document}
\newcommand{\Si}{\Sigma}
\newcommand{\tr}{{\rm tr}}
\newcommand{\ad}{{\rm ad}}
\newcommand{\Ad}{{\rm Ad}}
\newcommand{\ti}[1]{\tilde{#1}}
\newcommand{\om}{\omega}
\newcommand{\Om}{\Omega}
\newcommand{\de}{\delta}
\newcommand{\al}{\alpha}
\newcommand{\te}{\theta}
\newcommand{\vth}{\vartheta}
\newcommand{\be}{\beta}
\newcommand{\la}{\lambda}
\newcommand{\La}{\Lambda}
\newcommand{\D}{\Delta}
\newcommand{\ve}{\varepsilon}
\newcommand{\ep}{\epsilon}
\newcommand{\vf}{\varphi}
\newcommand{\vfh}{\varphi^\hbar}
\newcommand{\vfe}{\varphi^\eta}
\newcommand{\fh}{\phi^\hbar}
\newcommand{\fe}{\phi^\eta}
\newcommand{\G}{\Gamma}
\newcommand{\ka}{\kappa}
\newcommand{\ip}{\hat{\upsilon}}
\newcommand{\Ip}{\hat{\Upsilon}}
\newcommand{\ga}{\gamma}
\newcommand{\ze}{\zeta}
\newcommand{\si}{\sigma}

\def\clA{\mathcal{A}}
\def\clF{\mathcal{F}}
\def\clG{\mathcal{G}}
\def\clH{\mathcal{H}}
\def\clM{\mathcal{M}}
\def\clR{\mathcal{R}}
\def\clU{\mathcal{U}}
\def\clO{\mathcal{O}}
\def\clL{\mathcal{L}}
\def\clN{\mathcal{N}}
\def\clS{\mathcal{S}}
\def\clT{\mathcal{T}}
\def\clY{\mathcal{Y}}
\def\clW{\mathcal{W}}
\def\clZ{\mathcal{Z}}

\def\bfa{{\bf a}}
\def\bfb{{\bf b}}
\def\bfc{{\bf c}}
\def\bfd{{\bf d}}
\def\bfe{{\bf e}}
\def\bff{{\bf f}}
\def\bfm{{\bf m}}
\def\bfn{{\bf n}}
\def\bfp{{\bf p}}
\def\bfu{{\bf u}}
\def\bfv{{\bf v}}
\def\bft{{\bf t}}
\def\bfx{{\bf x}}
\def\bfg{{\bf g}}
\def\bfC{{\bf C}}
\def\bfA{{\bf A}}
\def\bfS{{\bf S}}
\def\bfJ{{\bf J}}
\def\bfI{{\bf I}}
\def\bfP{{\bf P}}
\def\bfr{{\bf r}}
\def\bfU{{\bf U}}
\def\bfPhi{{\bf \Phi}}
\def\bfphi{{\bf \phi}}

\def\bfal{\breve{\al}}
\def\bfbe{\breve{\be}}
\def\bfga{\breve{\ga}}
\def\bfnu{\breve{\nu}}
\def\bfsi{\breve{\sigma}}

\def\bPhi{\bar{\Phi}}

\def\hS{{\hat{S}}}
\def\ad{\rm ad}
\def\Ad{\rm Ad}

\newcommand{\li}{\lim_{n\rightarrow \infty}}
\def\mapright#1{\smash{
\mathop{\longrightarrow}\limits^{#1}}}

\newcommand{\mat}[4]{\left(\begin{array}{cc}{#1}&{#2}\\{#3}&{#4}
\end{array}\right)}
\newcommand{\thmat}[9]{\left(
\begin{array}{ccc}{#1}&{#2}&{#3}\\{#4}&{#5}&{#6}\\
{#7}&{#8}&{#9}
\end{array}\right)}

\newcommand{\thch}[4]{\theta\left[
\begin{array}{c}{#1}\\{#2}
\end{array}
\right]({#3};{#4})}

\newcommand{\beq}[1]{\begin{equation}\label{#1}}
\newcommand{\eq}{\end{equation}}
\newcommand{\beqn}[1]{\begin{eqnarray}\label{#1}}
\newcommand{\eqn}{\end{eqnarray}}
\newcommand{\p}{\partial}
\newcommand{\di}{{\rm diag}}
\newcommand{\oh}{\frac{1}{2}}
\newcommand{\su}{{\bf su_2}}
\newcommand{\uo}{{\bf u_1}}
\newcommand{\SL}{{\rm SL}(2,{\mathbb C})}
\newcommand{\GLN}{{\rm GL}(N,{\mathbb C})}
\newcommand{\PGLN}{{\rm PGL}(N,{\mathbb C})}

\def\SUN{{\rm SU}(N)}
\def\sun{{\rm su}(N)}
\def\sln{{\rm sl}(N, {\mathbb C})}
\def\sl2{{\rm sl}(2, {\mathbb C})}
\def\SLN{{\rm SL}(N, {\mathbb C})}
\def\SLT{{\rm SL}(2, {\mathbb C})}
\newcommand{\gln}{{\rm gl}(N, {\mathbb C})}
\newcommand{\PSL}{{\rm PSL}_2( {\mathbb Z})}
\def\f1#1{\frac{1}{#1}}
\def\lb{\lfloor}
\def\rb{\rfloor}
\def\sn{{\rm sn}}
\def\cn{{\rm cn}}
\def\dn{{\rm dn}}
\newcommand{\rar}{\rightarrow}
\newcommand{\upar}{\uparrow}
\newcommand{\sm}{\setminus}
\newcommand{\ms}{\mapsto}
\newcommand{\bp}{\bar{\partial}}
\newcommand{\bz}{\bar{z}}
\newcommand{\bw}{\bar{w}}
\newcommand{\bA}{\bar{A}}
\newcommand{\bL}{\bar{L}}
\newcommand{\btau}{\bar{\tau}}

\newcommand{\Sh}{\hat{S}}
\newcommand{\vtb}{\theta_{2}}
\newcommand{\vtc}{\theta_{3}}
\newcommand{\vtd}{\theta_{4}}

\def\mC{{\mathbb C}}
\def\mZ{{\mathbb Z}}
\def\mR{{\mathbb R}}
\def\mN{{\mathbb N}}
\def\mP{{\mathbb P}}

\def\frak{\mathfrak}
\def\gg{{\frak g}}
\def\gJ{{\frak J}}
\def\gS{{\frak S}}
\def\gL{{\frak L}}
\def\gG{{\frak G}}
\def\gk{{\frak k}}
\def\gK{{\frak K}}
\def\gl{{\frak l}}
\def\gh{{\frak h}}
\def\gH{{\frak H}}

\newcommand{\ran}{\rangle}
\newcommand{\lan}{\langle}
\def\f1#1{\frac{1}{#1}}
\def\lb{\lfloor}
\def\rb{\rfloor}
\newcommand{\slim}[2]{\sum\limits_{#1}^{#2}}

%\def\theequation{\thesubsection.\arabic{equation}}% the equation
               % number now does not include the section number;
               % \setcounter{equation}{0} should be put after every
               % \section{} command!!!
\newcommand{\sect}[1]{\setcounter{equation}{0}\section{#1}}
\renewcommand{\theequation}{\thesection.\arabic{equation}}
\newtheorem{predl}{Proposition}[section]
\newtheorem{defi}{Definition}[section]
\newtheorem{rem}{Remark}[section]
\newtheorem{cor}{Corollary}[section]
\newtheorem{lem}{Lemma}[section]
\newtheorem{theor}{Theorem}[section]

%\vspace{0.3in}

%\vspace{0.3in}
\begin{flushright}
% ITP-UH-01/06\\
 ITEP-TH-08/08
\end{flushright}
\vspace{10mm}
\begin{center}
{\Large{\bf Three lectures on classical integrable systems and gauge field theories}}
\footnote{Lectures given at JINR (Dubna, March 2007) and
Advanced Summer School on Integrable Systems and Quantum Symmetries
(Prague, June, 2007)}

\vspace{5mm}

 M.Olshanetsky,
\footnote{The work was supported by grants RFBR-06-02-17382, RFBR-06-01-92054-CE$_a$,
NSh-8065-2006.2.}
\\
\emph{e-mail olshanet@itep.ru}
\vspace{3mm}

{\it
 Institute of Theoretical and Experimental Physics, Moscow,}

\vspace{5mm}
\end{center}

\begin{abstract}
In these lectures I consider the Hitchin integrable systems and their
relations with the self-duality equations and the twisted super-symmetric Yang-Mills
theory in four dimension follow Hitchin and Kapustin-Witten. I define the Symplectic Hecke correspondence between
different integrable systems. As an example I consider Elliptic Calogero-Moser
system and integrable Euler-Arnold top on coadjoint orbits of the group GL(N,C)
and explain the Symplectic Hecke correspondence for these systems.
\end{abstract}
%\today
\tableofcontents

\part{Lecture 1}
\section{Introduction}

Some interrelations between  classical
 integrable systems and field theories in dimensions 3 and 4  were proposed by N.Hitchin twenty years ago \cite{H1,H2}. This approach to integrable systems has some advantages. It immediately leads to the
Lax representation with a spectral parameter, allows to prove in some cases the algebraic integrability
and to find separated variables \cite{Hu,GNR}. It was found later that
some well-known integrable systems can be derived in this way \cite{Ma,GN,ER,LO,Ne,Kr1,LOZ1,LOZ2}.

It was demonstrated in \cite{GKMMM} that there exists an integrable regime
in $\clN=2$ supersymmetric Yang-Mills theory in four dimension, which described by Sieberg and Witten \cite{SW}. A general picture  of interrelations between integrable models and gauge theories in dimensions 4,5 and 6 was presented in review \cite{GM}.

Some new aspects of interrelations between integrable systems and gauge theories
 were found recently in the framework of four-dimensional
reformulation of the geometric Langlands program  \cite{KW,GW,W}. These lectures  take into account
this approach, but also is based on the papers and reviews \cite{H1,H2,LOZ1,LOZ2,OP,Do1,Et,Zo}.

The derivation of integrable systems from field theories is based on the symplectic or the
Poisson reduction. This construction is familiar in gauge field theories.
The physical degrees of freedom in gauge theories are defined upon imposing the first and second the class
constraints. The first class constraints are analogs of the Gauss law generating
the gauge transformations. A combination of the Gauss law and constraints coming from the
gauge fixing yields second class constraints.

 We start with gauge theories that have some important properties. First, they have
 at least a finite number of independent  conserved
quantities. After the reduction they will play the role of integrals of motion. Next, we assume that after a gauge fixing and solving the
 constraints, the reduced phase space becomes a finite-dimensional manifold
 and its  dimension  is twice of number of integrals.
The latter property provides the complete integrability.
It is, for example, the theory of the Higgs bundles describing the Hitchin integrable systems \cite{H1}.
This theory corresponds to a gauge theory in three dimension. On the other hand, the similar type
of constraints arises in reduction of the self-duality equations in the four-dimensional Yang-Mills
theory \cite{H1}, and in the fourdimensional
$\clN=4$ supersymmetric Yang-Mills theory\cite{KW}  after reducing them to a space of dimension two.

We also analyze the problem of the classification of integrable systems.
Roughly speaking two integrable  systems are called equivalent,
if the original field theories are gauged equivalent. We extend the gauge transformations by allowing singular
gauge transformations of a special kind. On the field theory side these transformations
corresponds to monopole configurations, or, equivalently, the including
 the t'Hooft operators \cite{tH,Ka}. For some particular
examples we establish in this way
an equivalence of integrable systems of particles (the Calogero-Moser
systems) and integrable Euler-Arnold tops. It turns out that this equivalence
is the same as equivalence of two types $R$-matrices of dynamical  and
vertex type \cite{B,JMO}.

Before consider concrete cases we remind the main definitions
of completely integrable systems \cite{Do1,Et,Ar}.

%%%%%%%%%%%%%%%%%%%%%%%%%%%%%%%%%%%%%%%%%%%%%%%%%%%%%%%%%%%%%%%%%%%%%%%%%%%%%%%%%%%%%%%%%%%%%%
%%%%%%%%%%%%%%%%%%%%%%%%%%%%%%%%%%%%%%%%%%%%%%%%%%%%%%%%%%%%%%%%%%%%%%%%%%%%%%%%%%%%%%%%%%%%%%

\section{Classical Integrable systems}
\setcounter{equation}{0}

Consider a smooth symplectic manifold $\clR\,$ of $\dim(\mathcal{R})=2l$.
It means that there exists a closed non-degenerate  two-form $\om$, and the inverse
bivector $\pi\,$
$\,(\om_{a,b}\pi^{bc}=\de^c_a)$, such that the space $C^\infty(\clR)$ becomes a Lie
algebra (the Poisson algebra) with respect to the Poisson brackets
$$
\{F,G\}=\lan dF|\pi| dG\ran=\p_aF\pi^{ab}\p_bG
$$
Any $H\in C^\infty(\clR)$ defines a Hamiltonian vector field on $\clR\,$
$$
H\to\lan dH|\pi=\p_aH\pi^{ab}\p_b=\{H,\,\}\,.
$$
A Hamiltonian system is a triple $(\clR,\pi,H)$ with the Hamiltonian flow
$$
\p_tx^a=\{H,x^a\}=\p_bH\pi^{ba}\,.
$$
A Hamiltonian system is called {\it completely integrable}, if it satisfies the
following conditions
\begin{itemize}
  \item there exists $l$ Poisson commuting
Hamiltonians on $\clR$ ({\it integrals of motion})
$I_1,\ldots,I_l$
  \item Since the integrals commute the set
  $T_c=\{\{I_j=c_j\}$ is invariant with respect to the Hamiltonian flows
  $\{I_j,~\}$. Then being restricted on $T_c\,$, $\,I_j(x)$ are functionally independent  almost for all $x\in T_c$, i.e. $\det(\p_aI_b)(x)\neq 0$.
\end{itemize}

In this way we come to the hierarchy of commuting flows on $\clR$
\beq{hh}
\p_{t_j}\bfx=\{I_j(\bfx),\bfx\}\,.
\eq
$T_c=$ is a submanifold $T_c\subset\clR$.
It is is a Lagrangian submanifold , i.e. $\om$ vanishes on
$T_c$.
If $T_c$ is compact and connected, then it is diffeomorphic to a $l$-dimensional torus.
The torus $T_c$ is called {\it the Liouville torus}. In a neighborhood of $T_c$ there is a projection
 \beq{pro}
 p\,:\,\clR\to B\,,
 \eq
  where the Liouville tori are generic fibers and the base of fibration $B$ is parameterized by the values of integrals.
 The coordinates on a Liouville torus ("the angle" variables) along with dual variables on $B$
("the action" variables) describe a linearized motion on the torus.
Globally, the picture can be more complicated. For some values of $c_j\,$ $\,T_c$ ceases to be
a submanifold. In this way the action-angle variables are local.

%Due to the Liouville theorem these equations can be
%solved in quadratures.

% For generic values of the constants $c_j\,$ $\,\dim B=l$.
%A fiber $T$ of the

Here we consider a complex analog of this picture.
We assume that $\clR$ is a complex algebraic manifold and the symplectic
form $\om$ is a $(2,0)$ form, i.e. locally in the coordinates $(z^1,\bz^1,\dots,z^l,\bz^l)\,$
the form is represented as $\,\om=\om_{a,b}dz^a\wedge dz^b$. General fibers of
(\ref{pro})  are {\it abelian subvarieties} of $\clR$, i.e. they
are complex tori $\mC^l/\Lambda$, where the lattice $\Lambda$  satisfies the Riemann conditions.
Integrable systems in this situation are called {\it algebraically
integrable systems}.

\bigskip

Let two integrable systems are described
by  two isomorphic sets of the action-angle variables.
In this case the integrable systems can be considered as equivalent. Establishing equivalence in
terms of angle-action variables is troublesome. There exists a more direct way based on
{\it the Lax representation.} The Lax representation is one of the commonly accepted methods
to construct and  investigate integrable systems. Let $L(x,z),\,$ $\,M_1(x,z),\ldots,M_l(x,z)$
be a set of $l+1$ matrices
depending on $x\in\clR$ with a meromorphic dependence on
the {\it spectral parameter} $z\in\Si$, where $\Si$ is a Riemann surface.
\footnote{It will be explained below that $L$ and $M$ are sections of some vector bundles over $\Si$.}
It is called {\it a basic spectral curve}.
 Assume that
the commuting flows (\ref{hh}) can be rewritten in the matrix form
\beq{Lax}
\p_{t_j}L(\bfx,z)=[L(\bfx,z),M_j(\bfx,z)]\,.
\eq
Let $f$ be a non-degenerate matrix of the same order as $L$ and $M$.
The transformations
\beq{gau}
L'=f^{-1}Lf\,,~~M_j'=f^{-1}\p_{t_j}f+f^{-1}M_jf\,.
\eq
is called the gauge transformation because it preserves the Lax form (\ref{Lax}).
The flows (\ref{Lax}) can be considered as special gauge transformations
$$
L(t_1,\dots,t_l)=f^{-1}(t_1,\dots,t_l)L_0f(t_1,\dots,t_l)\,,
$$
where $L_0$ is independent on times and defines an initial data, and $M_j=f^{-1}\p_{t_j}f$.
Moreover, it follows from this representation that the quantities $\tr (L(\bfx,z))^j$
 are preserved by the flows and thereby can produce, in principle, the integrals of motion.
 As we mentioned above,
it is  reasonable to consider two integrable systems to be equivalent if their Lax matrices
are related by non-degenerate gauge transformation.

We relax the definition of the gauge transformations and
and assume that $\det f$ can have poles and zeroes on the basic spectral curve $\Si$ with some additional restrictions on $f$. This equivalence is called
{\it the Symplectic Hecke Correspondence}.
This extension of equivalence will be considered in these lectures in details.
The following systems are equivalent in this sense :\\
EXAMPLES
\begin{itemize}
          \item 1. Elliptic Calogero-Moser system $\Leftrightarrow$
Elliptic $\GLN$ Top\,, ~\cite{LOZ1} ;
          \item 2. Calogero-Moser field theory $\Leftrightarrow$
Landau-Lifshitz equation\,, ~\cite{LOZ1,Kr1};
          \item 3. Painlev\'{e} VI $\Leftrightarrow$ Zhukovsky-Volterra gyrostat\,, ~\cite{LOZ2}.
\end{itemize}
The first example will be considered  in Section 4.

\bigskip

The gauge invariance of the Lax matrices allows one to define the spectral curve
\beq{spc}
\mathcal{C}=\{(\la\in\mC\,,z\in\Si)\,|\,\det(\la-L(x,z))=0\}\,.
\eq
The Jacobian of $\mathcal{C}$ is an abelian variety of dimension $g$, where $g$ is the genus of $\mathcal{C}$.
If  $g=l=\oh\dim\,\clR$ then $\mathcal{J}$  plays the role of the Liouville torus
and the system is algebraically integrable. In generic cases $g>l$ and to prove the algebraic
integrability one should find additional reductions of the Jacobians, leading to abelian spaces
of dimension $l$.

\bigskip

 Finally we formulate two goals of these lectures
\begin{itemize}
  \item derivation of the Lax equation and the Lax matrices
  from a gauge theory;
  \item  explanation of the equivalence between integrable models
  by inserting t'Hooft operators in a gauge theory.
\end{itemize}

%%%%%%%%%%%%%%%%%%%%%%%%%%%%%%%%%%%%%%%%%%%%%%%%%%%%%%%%%%%%%%%%%%%%%%%%

\section{1d Field theory}
\setcounter{equation}{0}
The simplest integrable models such as the rational Calogero-Moser system,
the Sutherland model, the open Toda model can be derived from matrix models of a finite
order. Here we consider a particular case - the rational  Calogero-Moser system (RCMS) \cite{Ca,Mo}.

\subsection{Rational Calogero-Moser System (RCMS)}

The phase space  of the RCMS is
$$
\mathcal{R}^{RCM}=\mC^{2N}=\{(\bfv,\bfu)\}\,,~~
\bfv=(v_1,\dots v_N)\,,~\bfu=(u_1,\ldots u_N)
$$
with the canonical symplectic form
\beq{OM}
\om^{RCM}=\sum_{j=1}^Ndv_j\wedge du_j\,,~~
\{v_j,u_k\}=\de_{jk}\,.
\eq
The Hamiltonian describes interacting particles with complex coordinates
$\bfu=(u_1,\ldots u_N)$ and complex momenta $\bfv=(v_1,\dots v_N)$
$$
H^{RCM}=\oh\sum_{j=1}^Nv_j^2+\nu^2\sum_{j>k}\f1{(u_j-u_k)^2}\,.
$$
The Hamiltonian leads to the equations of motion
\beq{2.2}
\p_tu_j=v_j\,,
\eq
\beq{2.3}
\p_tv_j=-\nu^2\sum_{j>k}\f1{(u_j-u_k)^3}\,.
\eq

The equations of motion can be put in
the Lax form
\beq{2.1}
\p_t L(\bfv,\bfu)=[L(\bfv,\bfu),M(\bfv,\bfu)]\,.
\eq
Here $L,M$ are the $N\times N$ matricies of the form
$$
L=P+X\,,~~M=D+Y\,,
$$
\beq{2.1a}
P=\di(v_1,\dots v_N)\,,~~X_{jk}=\nu(u_j-u_k)^{-1}\,,
\eq
\beq{2.1b}
Y_{jk}=-\nu(u_j-u_k)^{-2}\,,~~ D=\di(d_1,\ldots d_N)\,,
\eq
$$
d_j=\nu\sum_{k\neq j}(u_j-u_k)^{-2}\,.
$$
The diagonal part of the Lax equation (\ref{2.1}) implies
$\p_tP=[X,Y]_{diag}$.
It coincides with (\ref{2.3}). The non-diagonal part
has the form
$$
\p_tX=[P,Y]+\left([X,Y]_{nondiag}-[X,D]\right)\,.
$$
It can be found that $[X,Y]_{nondiag}=[X,D]$ and  the equation
$\p_tX=[P,Y]$
coincides with (\ref{2.2}).

The Lax equations produces the integrals of motion
\beq{2.12a}
I_m=\f1{m}\tr(L^m)\,,~~\p_t\tr(L^m)=0\,,~~m=1,2,\ldots N\,.
\eq
It will be proved later that they are in involution
$\{I_m,I_n\}=0$.
In particular, $I_2=H^{RCM}$.
Eventually, we come to the RCMS hierarchy
\beq{2.12}
\p_jf(\bfv,\bfu)=\{I_j,f(\bfv,\bfu)\}\,.
\eq

%%%%%%%%%%%%%%%%%%%%%%%%%%%%%%%%%%%%%%%%%%%%%%%%%%%%%%%%%%%%%%%%%%%%%%%%

\subsection{Matrix mechanics and the RCMS}

This construction was proposed in Ref.\,\cite{OP1,KKS}.
Consider a  matrix model with the phase space   $\clR=\gln\oplus\gln$
$$
\mathcal{R}=(\Phi,\bA)\,,
~~\Phi\,,\bA\in\gln\,,\footnote{These notations will be justified in next Section}
$$
$$
\dim\mathcal{R}=2N^2\,.
$$
The symplectic form on $\mathcal{R}$ is
\beq{om}
\om=\tr(d\Phi\wedge d\bA)=\sum_{j,k}d\Phi_{jk}\wedge d\bA_{kj}\,.
\eq
The corresponding Poisson brackets have the form
$$
\{\Phi_{jk},\bA_{il}\}=\de_{ki}\de_{jl}\,.
$$
Choose $N$ commuting integrals
$$
I_m=\f1{m}\tr(\Phi^m)\,,~\{I_m,I_n\}=0\,,~m=1,\dots N
$$
Take as a Hamiltonian $H=I_2$. Then we come to the free motion on $\clR$
\beq{2.7}
\p_t\Phi=\{H,\Phi\}=0\,,
\eq
\beq{2.8}
\p_t\bA=\{H,\bA\}=\Phi\,.
\eq
Generally, we have a free matrix hierarchy
\beq{2.11}
\p_j\Phi=0\,,~\p_j\bA=\Phi^{j-1}\,,~~(\p_j=\{I_j,\,\})\,.
\eq
%%%%%%%%%%%%%%%%%%%%%%%%%%%%%%%%%%%%%%%%%%%%%%%%%%%%%%%%%%%%%%%%%%%%%%%%%%%%%%%%%%

\subsubsection{Hamiltonian reduction}

The form $\om$ and the the integrals $I_m$ are invariant with resect to the
action of the gauge group
$$
\mathcal{G}=\GLN\,,
$$
$$
\Phi\to f^{-1}\Phi f\,,~~\bA\to  f^{-1}\bA f\,,~~f\in\GLN\,.
$$
The action of gauge Lie algebra $Lie(\clG)=\gln$ is represented by the vector fields
\beq{gs}
V_\ep\Phi=[\Phi,\ep]\,,~~V_\ep\bA=[\bA,\ep]\,.
\eq
Let $\imath_\ep$ be the contraction operator with respect to the vector field $V_\ep\,$
$\,(\imath_\ep=\sum_{j,k}(V_\ep)_{jk}\frac{\p}{\p_{jk}})$ and
$\mathcal{L}_\ep=d\imath_\ep+\imath_\ep d$ is the
corresponding Lie derivative.
The invariance of the symplectic form and the integrals means that
$$
\mathcal{L}_\ep\om=0\,,~~\mathcal{L}_\ep I_m=0\,.
$$
Since the symplectic form is closed $d\om=0$, we have $d\imath_\ep\om=0$.
Then on the affine space  $\mathcal{R}\,$ the one-form $\imath_\ep\om$ is  exact
\beq{mom}
\imath_\ep\om=dF(\Phi,\bA,\ep)\,.
\eq
The function $F(\Phi,\bA,\ep)$ is called  {\it the momentum Hamiltonian}.
The Poisson brackets with the momentum Hamiltonian generate the gauge transformations:
$$
\{F,f(\Phi,\bA\}=\clL_\ep f(\Phi,\bA)\,.
$$
 The explicit form of the momentum Hamiltonian is
$$
F(\Phi,\bA,\ep)=\tr(\ep[\Phi,\bA])\,.
$$

Define {\it the moment map}
$$
\mu\,:\,\clR\to Lie^*(gauge~group)\sim \gln\,,
$$
$$
\mu(\Phi,\bA)=[\Phi,\bA]\,,~~(\Phi,\bA)\mapsto [\Phi,\bA]\,.
$$
Let us fix its value as
\beq{2.5}
\mu=[\Phi,\bA]=\nu J\,,
\eq
\beq{J}
J=\left(
                      \begin{array}{ccccc}
                        0 & 1 & \cdots &\cdots & 1 \\
                        1 & 0 & 1    & \cdots   & 1    \\
                        \vdots & \ddots & \ddots&\cdots & \vdots \\
                        1 & \cdots & \cdots & 1   & 0\\
                      \end{array}
                    \right)\,.
\eq
It follows from the definition of the moment map that
 (\ref{2.5}) is the first class constraints.
In particular, $\{F(\Phi,\bA,\ep),F(\Phi,\bA,\ep')\}=F(\Phi,\bA,[\ep,\ep'])$.
Note, that the matrix $J$ is degenerate and is conjugated to the diagonal matrix
$\di(N-1,-1,\dots -1)$.
Let $\clG_0$ be a subgroup of the gauge group preserving the moment value
$$
\mathcal{G}_0=\{f\in\mathcal{G}\,|\,f^{-1}Jf=J\}\,,~~(\dim(\mathcal{G}_0)=(N-1)^2+1)\,.
$$
In other words $\clG_0$ preserves the surface in $\clR$
\beq{F}
F^{-1}(\nu J)=\{[\Phi,\bA]=\nu J\}.
\eq
Let us fix a gauge on this surface with respect to the $\clG_0$ action.
It can be proved that generic matrices $\bA$ can be diagonalized
by  $\clG_0$
\beq{2.6}
f^{-1}\bA f=\bfu=\di(u_1,\ldots u_n)\,,~ f\in\clG_0\,.
\eq
In other words,  we have two conditions - the the first class constraints
(\ref{2.5}) and the gauge fixing (\ref{2.6}).
The reduced phase space $\clR^{red}$ is result of the putting the both
types of constraints
$$
\mathcal{R}^{red}=\mathcal{R}//\mathcal{G}=F^{-1}(\nu J)/\clG_0\,.
$$
It has dimension
$$
\left\{
\begin{array}{llll}
\dim(\mathcal{R}^{red})&=\dim(\mathcal{R})&-\dim(\mathcal{G})&-
\dim(\mathcal{G}_0)\,,\\
2N-2&=2N^2&-N^2&-(N-1)^2+1)
\end{array}
\right.
$$

Let us prove that $\mathcal{R}^{red}=\clR^{RCM}$ and that the
hierarchy (\ref{2.11}) being restricted on $\clR^{RCM}$ coincides
with the RCMS hierarchy (\ref{2.12}).
Let $f\in\clG_0$ diagonalizes  $\bA$ in (\ref{2.6}). Define
\beq{2.10}
L=f^{-1}\Phi f.
\eq
Then it follows from (\ref{2.7}) that $L$ satisfies the Lax equation
$$
\p_t\Phi=0\Rightarrow
\p_t L=[L,M]\,,~~(M=-f^{-1}\p_tf)\,.
$$
The moment constraints (\ref{F})
allows one to find the off-diagonal part of $L$. Evidently, it coincides
with $X$ (\ref{2.1a}).
The diagonal elements of $L$ are free parameters.
In a similar way the off-diagonal part $Y$ (\ref{2.1b}) of $M$
can be derived from the equation of motion for $\bA$  (\ref{2.8}).
Thereby,  we come to the Lax form of the equations of motion for RCMS.
Since $\Phi\to L$ and $\bA\to\bfu$, the symplectic form
 $\om$ (\ref{om}) coincides on $\clR^{RCM}$ with $\om^{RCM}$ (\ref{OM}).
It follows from (\ref{2.10}) that the integrals (\ref{2.12a}) Poisson commute.
Therefore, we obtain to the RCMS hierarchy.

The same system can be derive starting with the matrix mechanics
based on $\SLN$. In this case $I_1=\tr \Phi=0$ and thereby in the reduced system
$\sum v_j=0$.

%%%%%%%%%%%%%%%%%%%%%%%%%%%%%%%%%%%%%%%%%%%%%%%%%%%%%%%%%%%%%%%%%%%%%%%%
%%%%%%%%%%%%%%%%%%%%%%%%%%%%%%%%%%%%%%%%%%%%%%%%%%%%%%%%%%%%%%%%%%%%%%%%

\part{Lecture 2}
\section{3d field theory}
\setcounter{equation}{0}
\subsection{Hitchin systems}

{\bf 1.Fields}

Define a field theory on $(2+1)$ dimensional space-time of the form
$\mathbb{R}\times \Si_{g,n}$, where
$\Si_{g,n}$ is a Riemann surface of genus $g$ with a divisor $D=(x_1,\ldots,x_n)$ of $n$ marked points.
%$(z,\bz)$ - local complex coordinates.
\bigskip

\includegraphics{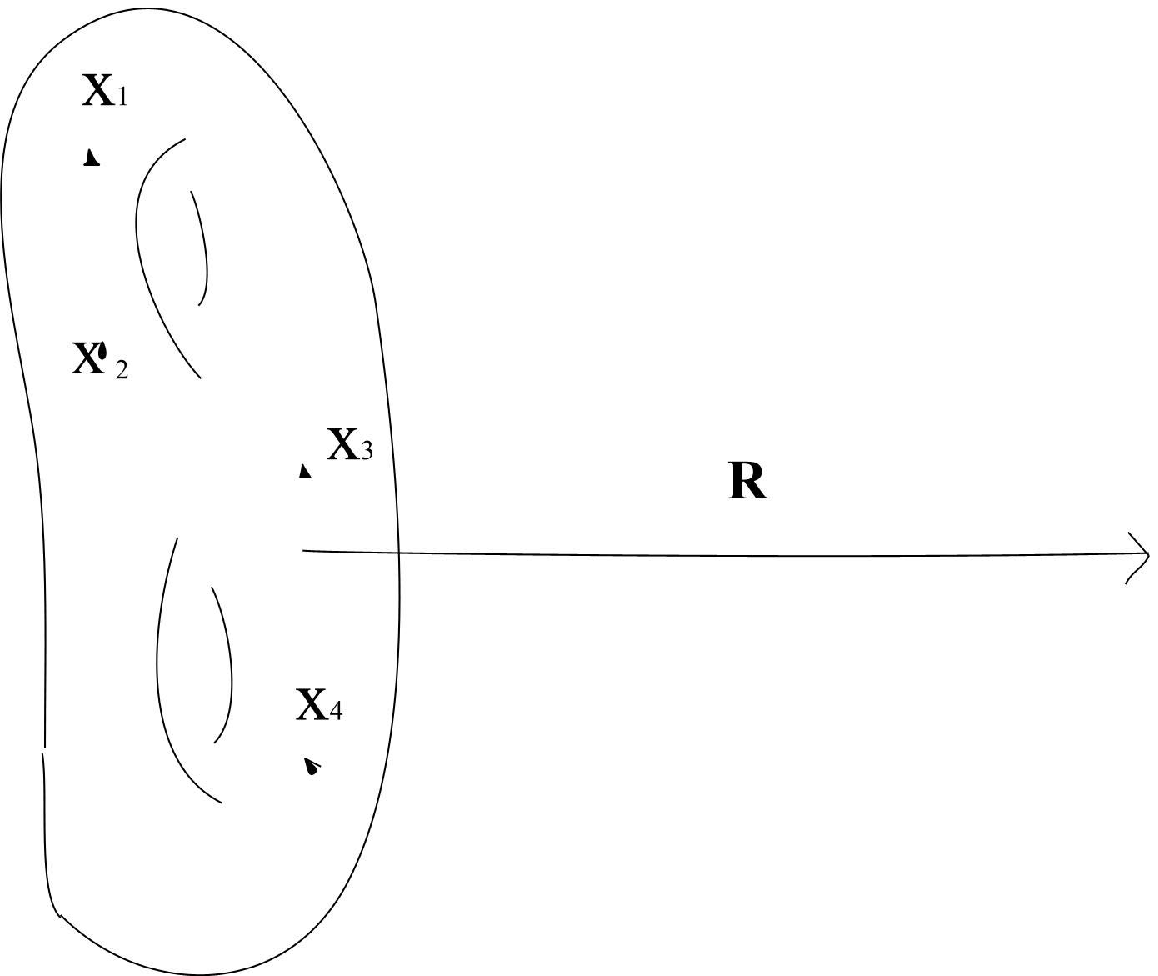}
$$
(x_1,\ldots,x_4)\, - ~marked~points\,.
$$

\bigskip

The phase space of the theory is defined by the following field content:

1) Consider a vector bundle $E$ of rank $N$ over $\Si_{g,n}$ equipped with the
connection $d''=\nabla_{\bz}\otimes d\bz$. It acts on the sections
$s^T=(s_1,\ldots,s_N)$ of $E$ as $d''s=\bp s+\bA s$. {\it The vector fields}
 $\bA(z,\bz)$ are $C^\infty$ maps $\Si_{g,n}\to \gln$.

2) The scalar fields ({\it the Higgs fields}) $\Phi(z,\bz)\otimes dz$,
$\Phi\,:\,\Si_{g,n}\to \gln$. The Higgs field is a section of the bundle
$\Om^{(1,0)}(\Si_{g,n}, End E)$.
It means that $\Phi$ acts on the sections $s_j\to \Phi_{kj}s_j\otimes dz$.
We assume that $\Phi$ has holomorphic poles at the
marked points $\Phi\sim\frac{\Phi^a}{z-x_a}+\dots\,$.\\
Let $(\al_1,\ldots,\al_g;\be_1,\ldots,\be_g)\,$ be a set of fundamental cycles of $\Si_{g,n}$,
$(\prod_j\al_j\be_j\al_j^{-1}\be_j^{-1}=1$). The bundle $E$ is defined by the
monodromy matrices $(Q_j,\La_j)$
$$
\al_j\,:\,s\to Q^{-1}_js\,,~~\be_j\,:\,\La^{-1}_js\,.
$$
Similarly, for $\bA$ and $\Phi$ we have
$$
\al_j\,:\,\bA\to Q_j\bp Q^{-1}_j+Q_j\bA Q^{-1}_j\,,~~
\be_j\,:\,\bA\to \La_j\bp \La^{-1}_j+\La_j\bA \La^{-1}_j\,
$$
\beq{Ph}
\al_j\,:\,\Phi\to Q_j\Phi Q^{-1}_j\,,,~~\be_j\,:\,\Phi\to \La_j\Phi \La^{-1}_j\,.
\eq

3)  {\it The spin variables} are attributed to the marked points $S^a\in\gln$, $a=1,\ldots,n$,\\
$S^a=g^{-1}S^a(0)g$, where $S^a(0)$ is a fixed element of $\gln$. In other
words, $S^a$ belong to  coadjoint orbits ${\cal O}^a$ of $\GLN$. They play the role
of non-abelian charges located at the marked points.

\bigskip

Let $\{T_\al\}\,$, $\,(\al=1,\dots,N^2)$  be a basis in the Lie algebra $\gln$, $[T_\al,T_\be]=C_{\al,\be}^\ga T_\ga$.
Define the Poisson structure on the space of fields:\\
1) The Darboux brackets for the fields $(A,\Phi)$:
$$
\bA(z,\bz)=\sum_\al \bA_{\al}(z,\bz) T_\al\,,~~\Phi(w,\bw)=\sum_\be \Phi_\be(w,\bw) T_\be\,.
$$
$$
\{\Phi_\al,(w,\bw),\bA_{\be}(z,\bz)\}=\lan T_\al T_\be\ran\de(z-w,w-\bw)\,,~~(\lan~~\ran=trace~in ~ad)
$$
2) Linear Lie brackets for the spin variables:\\
$\bfS^a=\sum_\al S^a_\al T_\al$
$$
\{S^a_\al,S^b_\be\}=\de^{a,b}C_{\al,\be}^\ga S_\ga^a\,.
$$
In this way we have defined the phase space
\beq{clr}
{\cal R}=(\bA,\Phi,\bfS^a)\,.
\eq
The  Poisson brackets are non-degenerate and the space $\clR$ is symplectic with the form
\beq{sympl}
\om=\om^0-\sum_{a=1}^n\int_{\Si_{g,n}}\om^a\de(z-x_a,\bz-\bar{x}_a)\,,
\eq
\beq{s1}
\om^0=\int_{\Si_{g,n}}\lan D\Phi\wedge D\bA\ran\,,
\eq
\beq{s2}
\om^a=\lan D(S^ag^{-1})\wedge Dg\ran\,.
\eq
The last form is the Kirillov-Kostant form on the coadjoint orbits. The fields
$(\Phi,\bA)$ are holomorphic coordinates on $\clR$
and the form $\om^0$ is the $(2,0)$-form in this complex structure on $\clR$.
Similarly, $(S^ag^{-1},g)$ are the holomorphic coordinates on the orbit $\clO^a$,
and $\om^a$ is also $(2,0)$ form.

{\bf 2. Hamiltonians}

 The traces $\lan \Phi^j\ran\,$ $\,(j=1,\ldots,N)$ of the Higgs field are
periodic  $(j,0)$-forms $\Omega^{(j,0)}(\Si_{g,n})$ with holomorphic poles of order $j$ at the marked points.
To construct integrals from $\lan \Phi^j\ran\,$ one should integrate them over $\Si_{g,n}$ and to this end prepare $(1,1)$-forms from the $(j,0)$-forms. For this purpose consider the space of smooth
$(1-j,1)$-differentials $\Omega^{(1-j,1)}(\Si_{g,n}\setminus D)$  vanishing at the marked points.
 Locally, they are represented as
 $\mu_j=\mu_j(z,\bz)\left(\frac{\p}{\p z}\right)^{j-1}\otimes d\bz$. In other words $\mu_j$ are $(0,1)$-forms taking values in degrees of vector fields $\clT$ on $\Si_{g,n}\setminus D$.
For example, $\mu_2$ is the Beltrami differential.

The product $\lan \Phi^j\ran\mu_j$ can be integrated over the surface. We explain below
that $\mu_j$ can be chosen as elements of basis in the cohomology space
$H^1(\Si_{g,n}\setminus D, \mathcal{T}^{\otimes j-1})$. This space has dimension
\beq{dime}
n_j=\dim H^1(\Si_{g,n},\clT^{\otimes(j-1)})=\left\{
\begin{array}{cc}
  (2j-1)(g-1)+jn\, & j>1 \\
  g & j=1
\end{array}
\right.
\eq
Let  $\mu_{j,k}$ be a basis in $H^1(\Si_{g,n},\clT^{\otimes(j-1)})\,$, $~(k=1,\ldots,n_j)$.
The product $\mu_{j,k}\lan\Phi^{j}\ran$ can be integrated to define the
 Hamiltonians
\beq{int}
I_{j,k}=\f1{j}\int_{\Si_{g,n}}
\mu_{j,k}\lan\Phi^{j}\ran\,,~~j=1,\ldots N\,.
\eq
%Let $\xi^{j,k_j}$ be a basis in $\Om^{(j,0)}(\Si_{g,n})$ dual to $\mu_{j,k}$
%$$
%\f1{j}\int_{\Si_{g,n}}\xi^{j,k_j}\mu_{m,i_m}=\de_{m}^j\de^{k_j}_{i_m}\,.
%$$
%Then the Hamiltonians are defined by the expansion
%$$
%\lan\Phi^{j}\ran=\sum_k^{n_j}H_{j,k}\xi^{j,k_j}\,.
%$$

It follows from (\ref{dime}) that
the number of the independent integrals $\sum n_j$ for $\GLN$ is
\beq{ni}
d_{N,g,n}=\sum_{j=1}^N n_j=(g-1)N^2+1+n\frac{N(N-1)}{2}\,.
\eq
Since $\lan\Phi\ran=0$ for $\SLN$ the number of the independent integrals is
\beq{ni1}
d_{N,g,n}=\sum_{j=2}^N n_j=(g-1)(N^2-1)+n\frac{N(N-1)}{2}\,.
\eq
The integrals $I_{(j,k)}$ are independent and
 Poisson commute
\beq{ii}
\{I_{(j_1,k_1)},I_{(j_2,k_2)}\}=0\,.
\eq
Thus we come to $d_{N,g,n}$ commuting flows on the phase space
${\cal R}(\bA,\Phi,\bfS^a)$
\beq{eq1}
\frac{\partial}{\partial t_{j,k}}\Phi=
\{\nabla I_{j,k},\Phi\}=0\,,
\eq
\beq{eq2}
\frac{\partial}{\partial t_{j,k}}\bA=\mu_{j,k}\Phi^{j-1}\,,
\eq
\beq{eq3}
\frac{\partial}{\partial t_{j,k}}\bfS^a=0\,.
\eq

{\bf 3. Action and gauge symmetries}

The same theory can be described by the action
$$
{\cal S}=\sum_{j=2}^N\sum_{k=1}^{n_j}\int_{\mR_{j,k}}\int_{\Si_{g,n}}
\left(\lan \Phi\p_{j,k}\bA\ran+
\sum_{a=1}^n\de(z-x_a,\bz-{\bar x}_a)\lan\bfS^ag_a^{-1}\p_{j,k}g_a\ran-I_{j,k}\right)dt_{j,k}\,,
$$
 where the time-like  Wilson lines at the marked points are included.

The action is gauge invariant with respect to the gauge group
$$
\clG_\mC=\{{\rm smooth~maps\,:~}\Si_{g,n}\to\GLN\}\,.
$$
The elements $f\in\clG_\mC$ are smooth and have the same monodromies as the Higgs field (\ref{Ph}).
%We put some restrictions on behavior of at the marked points. Let fix flags $Fl_a$ at
%$x_a\in D$. Assume that the gauge action preserves the flags. This property will be
%discussed later.

The action is invariant with respect to the gauge transformations
$$
\bA\to f^{-1}\bp f+f^{-1}\bA f\,,~~
\Phi\to f^{-1}\Phi f\,,
$$
$$
g_a\to g_af^a\,,~~\bfS^a\to (f^a)^{-1}\bfS^af^a\,,~~f^a=f(z,\bz)|_{z=x_a}\,.
$$
Consider the infinitesimal gauge transformations
$$
V_\ve \bA=\bp\ve+[ \bA,\ve]\,,~~V_\ve\Phi=[\Phi,\ve]\,,
$$
$$
V_\ve g_a=g_a\ve(x_a)\,,~~   V_\ve S^a=[S^a,\ve(x_a)]\,,~~~\ve\in\,Lie(\clG_\mC)\,.
$$
The Hamiltonian $F$ generating the gauge vector fields
$\imath_\ve\om=DF$ has the form
$$
F=\int_{\Si_{g,n}}\left\lan\ve(\bp\Phi+[\bA,\Phi]-
\sum_{a=1}^n\bfS^a\de(z-x_a,\bz-\bar{x}_a)
)\right\ran\,.
$$
The moment map
$$
\mu\,:\, {\cal R}(\bA,\Phi,\bfS^a)\to Lie^*(\clG_\mC)\,,~
\mu=\bp\Phi+[\bA,\Phi]-
\sum_{a=1}^n\bfS^a\de(z-x_a,\bz-\bar{x}_a)\,.
$$
The Gauss law  (the moment constraints) takes the form
\beq{HE}
\bp\Phi+[\bA,\Phi]=\sum_{a=1}^n\bfS^a\de(z-x_a,\bz-\bar{x}_a)\,.
\eq
Upon imposing these constraints the residues of the Higgs fields
become equal to the spin variables $Res \Phi_{z=x_a}=S^a$ in an analogy with the Yang-Mills
theory, where  the Higgs field corresponds to the electric field and $\bfS^a$ are analog of the electric charges.

The reduced phase space
$$
{\cal R}^{red}={\cal R}(\bA,\Phi,\bfS^a)/({\rm Gauss~ law})+({\rm gauge~fixing})
$$
defines the  physical degrees of freedom, and the reduced phase space is the
symplectic quotient
\beq{redu}
{\cal R}^{red}={\cal R}(\bA,\Phi,\bfS^a)//\clG_\mC\,.
\eq

\bigskip

%%%%%%%%%%%%%%%%%%%%%%%%%%%%%%%%%%%%%%%%%%%%%%%%%%%%%%%%%%%%%%%%%%%%%%%%%%%%%%%%%%%%%%

{\bf 4.Algebra-geometric approach}

The operator $d''$ acting on sections  defines a holomorphic structure on
the bundle $E$.
A section $s$ is holomorphic if
$$
(\bp+\bA)s=0\,.
$$
The moment constraint (\ref{HE}) means that the space of sections of the Higgs field
over $\Si_g\setminus D$ is holomorphic.

Consider the set of holomorphic structures $\clL=\{d_{\bA}\}$ on $E$.
Two holomorphic structure are called equivalent if the corresponding connections are
gauge equivalent. The moduli space of holomorphic structures is the quotient $\clL/\clG_\mC$.
Generically the quotient has very singular structure. To have a reasonable topology one should consider the so-called  stable bundles. The  stable bundles are generic and
 we consider the space of connection $\clL^{stable}$
 corresponding to the stable bundles. The quotient is called
{\it the moduli space of stable holomorphic bundles}
$$
\mathcal{M}(N,g,n)=\clL^{stable}/\mathcal{G}\,.
$$
It is a finite-dimensional manifold. The tangent space to  $\mathcal{M}(N,g,n)$ is isomorphic to\\
$H^1(\Si_{g,n},{\rm End}E)$. Its dimension can be extracted from
 the Riemann-Roch theorem and for curves without marked points $(n=0)$
$$
\dim H^0(\Si,{\rm End}E)-\dim H^1(\Si,{\rm End}E)=
(1-g)\dim G\,.
$$

%$$
%\dim(\mathcal{M}(N,g,n))=\dim T(\mathcal{M}(N,g,n))=\dim H^1(\Si_{g,n},{\rm End}E)\,.
%$$

%The sections of the tangent space $T\clL$   are  $(1,0)$-forms taking values in the %Lie algebra $\gln$
%$$
%T\clL=\Om^{(1,0)}(\Si,{\rm End}\,E)\,.
%$$

For stable bundles and $g>1\,$ $\,\dim (H^0(\Si,{\rm End}E))=1$ and
$$
\dim{\cal M}(N,g,0)=(g-1)N^2+1
$$
for $\GLN$, and
$$
\dim{\cal M}(N,g,0)=(g-1)(N^2-1)
$$
for $\SLN$.

Thus, in the absence of the marked points we should consider bundles
over curves of genus $g\geq 2$. But the curves of genus $g=0$ and $1$ are
important for applications to integrable systems.
Including  the marked points improves the situation.

We extend the moduli space by adding an additional data at the marked points.
Consider an  N-dimensional vector space $V$ and choose a flag
$Fl=(V_1\subset V_2\subset\ldots V_N=V)$. Note that  flag is a point in a homogeneous
 space called the flag variety  $Fl\in \GLN/B$, where $B$ is a Borel subgroup.
 If $(e_1,\ldots,e_N)$ is a basis in $V$ and $Fl$ is a flag
 $$
 Fl=\{V_1=\{a_{11}e_1\},\,V_2=\{a_{21}e_1+a_{22}e_2\},\ldots V_N=V\}
 $$
 then $B$ is the subgroup of lower triangular matrices. The flag variety
 has dimension $\oh  N(N-1)$.
    The moduli space ${\cal M}(N,g,n)$ is
 the moduli space  ${\cal M}(N,g,0)$ equipped with maps $g_a\in\GLN$ of $V$ to the fibers over the marked points $V\to E|_{x_a}$ , preserving $Fl$
in $V$. In other words
$g_a$ are defined up to the right multiplication of $B$ and therefore we supply the
moduli space ${\cal M}(N,g,0)$ with structure of the flag variety $\GLN/B$
at the marked points.
 We have a natural "forgetting" projection
$\pi\,:\,{\cal M}(N,g,n)\to{\cal M}(N,g,0)$. The fiber of this projection is the
product of $n$ copies of the flag varieties.
 The bundles with this structure are called {\it the quasi-parabolic
bundles}. The dimension of the moduli space of quasi-parabolic holomorphic bundles is
$$
\dim{\cal M}(N,g,n)=\dim{\cal M}(N,g,0)+
\oh n N(N-1)\,.
$$

For curves of genus  $g>1\,$ $\,\dim({\cal M}(N,g,n))$ is independent on
 degree of the bundles $d=deg(E)=c_1(\det E)$.
In fact, we have a disjoint union of
components labeled by the corresponding degrees of the bundles  \mbox{${\cal M}=\bigsqcup{\cal M}^{(d)}$.}
For elliptic curves $(g=1)$ one has
$$
\dim H^1(\Si,{\rm End}E)=\dim H^0(\Si,{\rm End}E),
$$
and $\dim H^0(\Si,{\rm End}E)$ does depend on deg$(E)$. Namely,
\beq{2.0}
\dim({\cal M}(N,1,0,d))= {\rm g.c.d.} (N,d)\,.
\eq
In this case the structure of the moduli space  for the trivial bundles
(i.e. with ${\rm deg}(E)=0$) and, for example, for bundles with
${\rm deg}(E)=1$ are different.

Now consider the Higgs field $\Phi$. As we already mentioned $\Phi$  defines an endomorphism of the bundle $E$
$$
\Phi\,:\Om^{(0)}(\Si_{g,n},E)\to\Om^{(1,0)}(\Si_{g,n},E)\,,~~s\to\Phi s\otimes dz\,.
$$
Similarly, they can be described as sections of $\Om_{C^\infty}^{0}(\Si_{g,n},{\rm End}\,E\otimes K_D)$. Here $K_D$ is the canonical class on $\Si\setminus D$
that locally apart from $D$ is represented as $dz$. Remind that $\Phi$ has poles at $D$. On the other hand, as it follows from the definition of the symplectic structure
   (\ref{s1}) on the set of pairs $(\Phi,\bA)$, the Higgs field plays the role of
a "covector" with  respect to the vector $\bA$. In this way
the Higgs field $\Phi$ is a section of the cotangent bundle $T^*\clL^{stable}$.

The pair of the holomorphic vector bundle and the Higgs field $(E,\Phi)$ is called
{\it the Higgs bundle}. The reduced phase space (\ref{redu}) is  {\it the moduli space of  the quasi-parabolic Higgs bundles}. It is the cotangent bundle
\beq{mhb}
\clR^{red}=T^*\clM(N,g,n,d)\,.
\eq

Due to the Gauss law  (\ref{HE}) the Higgs fields are holomorphic on $\Si\setminus D$. Then on the reduced space $\clR^{red}$
\beq{Phi}
\Phi\in H^0(\Si_{g,n},{\rm End}^*\,E\otimes K_D)\,.
\eq

A part of $T^*\clM(N,g,n,d)$ comes from the cotangent bundle to the flag varieties $T^*(G/B)_a$ located at the marked points. Without the null section $T^*(G/B)_a$ is isomorphic to
a unipotent coadjoint orbit, while the null section is the trivial orbit.
Generic coadjoint orbits passing through a semi-simple element of $\gln$ is an
affine space over  $T^*(G/B)_a$. In this way we come to the moduli space of the quasi-parabolic Higgs bundles \cite{Si}. It has
dimension
\beq{dime}
\dim{\cal R}^{red}=2N^2(g-1)+2+ N(N-1)n\,.
\eq
This formula is universal and valid also for $g=0,1$ and
does not depend on  deg$(E)$.
At the first glance, for $g=1$ this formula contradict to (\ref{2.0}). In fact,
we have a residual gauge symmetry generated by subgroup of the Cartan group of
$\GLN$. The symplectic reduction with respect to this symmetry kill these degrees
of freedom and we come to $\dim{\cal R}^{red}=2+ N(N-1)n$ (see (\ref{dime}).
We explain this mechanism on a particular example in Section {\bf 4.2.2}.
The formula (\ref{dime}) suggests that the phase spaces corresponding to
bundles of different degrees may be symplectomorphic. We will see soon that it is the case.

It follows from (\ref{Phi}) that $\lan\Phi^j\ran\in H^0(\Si_{g,n}, K_D^j)$.
In other words $\lan\Phi^j\ran$ are meromorphic forms on the curve with the
poles of order $j$ at the divisor $D$.
Let $\varsigma^{jk}$ be a basis of $H^0(\Si_{g,n}, K_D^j)$.
Then
\beq{integ}
\f1{j}\lan\Phi^j\ran=\sum_{k=1}^{n_j}I_{jk}\varsigma^{jk}\,.
\eq
The introduced above the basis $\mu_{jk}$ in $H^1(\Si_{g,n}\setminus D, \mathcal{T}^{\otimes j-1})$ is dual to the basis $\varsigma^{jk}$
$$
\int_{\Si_{g,n}}\mu_{jk}\varsigma^{lm}=\de_j^l\de_k^m\,.
$$
Then the coefficients of the expansion (\ref{integ}) coincide with the integrals
 (\ref{int}). The dimensions $n_j$  (\ref{dime}) can be calculated as
 $\dim H^0(\Si_{g,n}, K_D^j)$.

 The symplectic reduction
preserves the involutivity (\ref{ii}) of the integrals (\ref{int}).
Since
$$
\fbox{$
\oh\dim\,T^*\clM(N,g,n)=number~of~integrals
$}
$$
(see (\ref{ni}), (\ref{ni1})) we come to integrable systems on the moduli
space of the quasi-parabolic Higgs bundles $\clR^{red}$.

For $\GLN$ the Liouville torus is the Jacobian of  the spectral curve
$\mathcal{C}$ (\ref{spc}). Consider bundles with the structure group replaced by a reductive
group $G$.
 The algebraic integrability for
$g>1$ and $G$ is a classical simple group was proved in \cite{H1}. The case of exceptional
groups was considered in \cite{Fa,Do}.

\bigskip

%%%%%%%%%%%%%%%%%%%%%%%%%%%%%%%%%%%%%%%%%%%%%%%%%%%%%%%%%%%%%%%%%%%%%%%%%%%%%%%%%%%%%%%%%%%%%%

{\bf 5.Equations of motion on the reduced phase space}

Let us fix a gauge $\bA=\bA_{0}$. For an arbitrary connection $\bA$ define a gauge
transform
$$
f[\bA]\,:\,\bA\to \bA_{0}\,,~~~\bA_{0}=(f^{-1}\bp f)[\bA]+f^{-1}[\bA]\bA f[\bA]\,.
$$
Then $f[\bA]$ is an element of the coset space $\clG_\mC/\clG_0$, where the subgroup
$\clG_0$ preserves the gauge fixing
$$
\clG_0=\{\,f\,| \bp f+[\bA_{0},f]=0\}\,.
$$
The same gauge transformation brings the Higgs field to the form
$$
L=f^{-1}[\bA]\Phi f[\bA]\,.
$$
The equations of motion for $\Phi$ (\ref{eq1}) in terms of $L$  takes the form
of the Lax equation
\beq{lam}
\fbox{$ \p_{j,k}L=[L,M_{j,k}]$}\,,
\eq
where $M_{j,k}=f^{-1}[\bA]\p_{j,k}f[\bA] $.
Therefore, the Higgs field becomes after reduction the Lax matrix.
The equations (\ref{lam}) describes {\it the Hitchin integrable hierarchy}.

The matrix $M_{j,k}$ can be extracted from the second equation (\ref{eq2})
\beq{mam}
\bp M_{j,k}-[M_{j,k},\bA_{0}]=\p_{j,k}\bA_{0}-L^{j-1}\mu_{j,k}\,.
\eq
The Gauss law restricted on $\clR^{red}$ takes the form
\beq{gal}
\bp L+[\bA_{0},L]=\sum_{a=1}^n\bfS^a\de(x_a,\bar{x}_a)\,.
\eq
Thus, the Lax matrix is the matrix Green function of the operator
$\bp+\bA_{0}$ on $\Si_{g,n}$ acting in the space $\Om^{(1,0)}(\Si_{g,n},End\,E)$.

The linear system corresponding to the integrable hierarchy takes the following form.
Consider a section $\psi$ of the vector bundle $E$. The section is called
 the Baiker-Akhiezer function if it is a solution of the linear system for
$$
\left\{
\begin{array}{cl}
  1. & (\bp+\bA_{0})\psi=0\,, \\
  2. & (\la-L)\psi=0\,, \\
  3. & (\p_{j,k}+M_{j,k})\psi=0 \,.
\end{array}
\right.
$$
The first equation means that $\psi$ is a holomorphic section.
Compatibility of the first equation and the second equation is the Gauss law
(\ref{gal}) and the first equation and the last equation is the Lax equations
(\ref{lam}).

In term of the Lax matrix the integrals of motion $I_{jk}$
are expressed by the integrals (\ref{int})
\beq{in}
I_{jk}=\f1{j}\int_{\Si_{g,n}}\mu_{jk}\tr(L(x,z))^j\,,
\eq
or by the expansion (\ref{integ})
\beq{int2}
\f1{j}\lan L^j\ran=\sum_{k=1}^{n_j}I_{jk}\varsigma^{jk}\,.
\eq

The moduli space of the Higgs bundles (\ref{mhb}) is parameterized  by the pairs
$({\bf A}_0,L)$. The projection (\ref{pro})
$$
T^*\clM(N,g,n)\to B=\sum_{j=1}^N H^0(\Si_{g,n}\setminus D,K^j_D)
$$
is called {\it the Hitchin fibration}.

%%%%%%%%%%%%%%%%%%%%%%%%%%%%%%%%%%%%%%%%%%%%%%%%%%%%%%%%%%%%%%%%%%%%%%%%%%%%%%%%%%%%%%

An illustrative examples of the Hitchin construction is the Higgs  bundles over
elliptic curves. These cases will be described explicitly in next subsections.

\subsection{N-body Elliptic Calogero-Moser System (ECMS)}

{\bf 1.Description of system}

Let $C_\tau$ be an elliptic curve $\mC/(\mZ+\tau\mZ)$, $\,(Im\tau>0)$.
The phase space ${\cal R}^{ECM}\,$ of ECMS is described by $N$ complex coordinates
and their momenta
$$
\left\{
\begin{array}{ll}
\bfu=(u_1,\ldots,u_N)\,,& (u_j\in C_\tau)\, -\,{\rm coordinates~ of~ particles}\,,\\
\bfv=(v_1,\ldots,v_N)\,, & (v_j\in\mC)\, -\,{\rm momentum~vector}
\end{array}
\right.
$$
with the Poisson brackets $\{v_j,u_k\}=\de_{jk}$.
\bigskip

\includegraphics{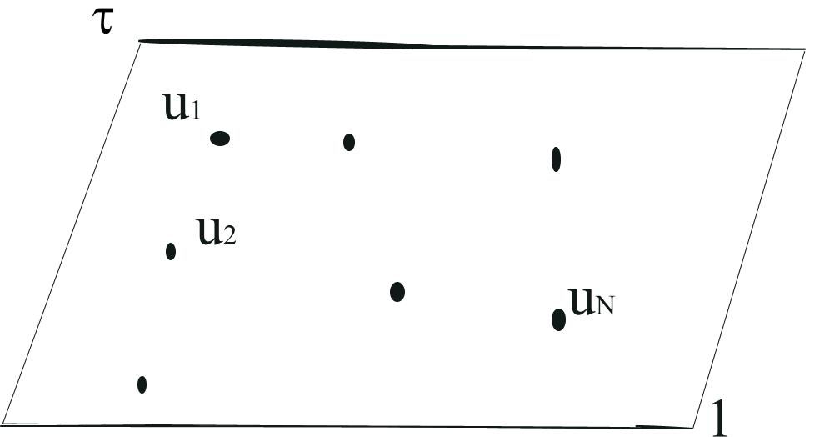}

\bigskip

The Hamiltonian takes the form
\beq{hcm}
H^{CM}=\oh|\bfv|^2+\nu^2\sum_{j<k}\wp(u_j-u_k)\,.
\eq
 Here
$\nu^2$ is a coupling constant and $\wp(z)$ - is the Weierschtrass function.
It is a double periodic meromorphic function $\wp(z+1)=\wp(z+\tau)=\wp(z),$ with a second order pole
$\wp(z)\sim z^{-2}\,$, $z\to 0$.

The system has the Lax representation \cite{Kr2} with the Lax matrix
\beq{lcm}
L^{CM}=iV+X\,,~~V=\di(v_1,\ldots,v_N)\,,
\eq
\beq{xjk}
X_{jk}=\nu\bfe(\frac{z-\bz}{\tau-\bar{\tau}}(u_j-u_k))\phi(u_j-u_k,z)\,,
~~\bfe(x)=\exp 2\pi ix\,,
\eq
where
\beq{phii}
\phi(u,z)=\frac{\theta(u+z)\theta'(0)}
{\theta(u)\theta(z)}\,,
\eq
and
\beq{thet}
\theta(z)=q^{\frac
{1}{8}}\sum_{n\in {\bf Z}}(-1)^n\exp 2\pi\imath(\oh n(n+1)\tau+nz)\,,~~
q=\exp 2\pi i\tau
\eq
is the standard theta-function with a simple zero at $z=0$ and the monodromies
\beq{mon}
\theta(z+1)=-\theta(z)\,,~~\theta(z+\tau)=-q^{-\oh}e^{-2\pi iz}\theta(z)\,.
\eq
Then from (\ref{mon}) that
\beq{phi}
\phi(u,z+1)=\phi(u,z)\,,~\phi(u,z+\tau)=\bfe(-u)\phi(u,z)\,.
\eq
and $\phi(u,z)$ has a simple pole at $z=0\,$
\beq{resi}
Res\,\phi(u,z)|_{z=0}=1\,.
\eq

%The off-diagonal part of matrix $M$ related to the Hamiltonian $H^{CM}$ (\ref{hcm}) can be %found from (\ref{mam}).
\bigskip

{\bf 2. ECMS and the Higgs bundles \cite{GN,ER}}

To describe the ECMS as the Hitchin system   consider a vector bundle $E$ of rank
$N$ and degree $0$ over an elliptic curve $\Si_{1,1}$ with one marked point. We assume that
the curve is isomorphic to $C_\tau =\mC/(\mZ+\tau\mZ)$.
The quasi-parabolic Higgs bundle $T^*E$ has coordinates
$$
\mathcal{R}^0=\{\Phi(z,\bz)\,,\,\bA(z,\bz)\,,\,S\}\,,~~\Phi,\bA\in\gln\,,~S\in{\cal O}\,,
$$
 where  ${\cal O}$ is a  degenerate orbit at the marked point $z=0$
 $$
 {\cal O}=\{S=g^{-1}S^0g\,|\,g\in\GLN\,,\,S^0=\nu J\}\,,
 $$
and $J$ is the matrix (\ref{J}). The orbit has dimension $\dim({\cal O})=2N-2$.

For degree zero bundles the monodromies around
the two fundamental cycles can be choosen as $Q_1=Id$ and
$\La_1=\bfe(\bfu)$, where $\bfe(\bfu)=\di(\exp 2\pi i u_1,\ldots,2\pi i u_N)$.
 A section with this monodromies is
\beq{sect}
s^T=(s_1,\ldots,s_N)\,,~~s_j=\phi(u_j,z)\,.
\eq
where $\phi(u_j,z)$ is  (\ref{phii}).
It follows from (\ref{phi}) that the section  has the prescribed monodromies.

For the fields and the gauge group we have the same monodromies
$$
\bA(z+1)=\bA(z)\,,~\Phi(z+1)=\Phi(z)\,,
$$
$$
\bA(z+\tau)=\bfe(\bfu)\bA(z)\bfe(-\bfu)\,,~
\Phi(z+\tau)=\bfe(\bfu)\Phi(z)\bfe(-\bfu)\,,
$$
$$
f(z+1,\bz+1)=f(z,\bz)\,,~
f(z+\tau,\bz+\bar{\tau})=\bfe(\bfu)f(z,\bz)\bfe(-\bfu)\,\,
$$

It can be proved that for  bundles of degree zero generic connections is trivial
$\bA=-\bp ff^{-1}$ and therefore
\beq{A0}
\bA\to\bA_{0}=0\,.
\eq
It means that stable bundles $E$ of rank $N$ are decomposed into the direct
sum of line bundles
$$
E=\oplus_{j=1}^N\clL_j\,,
$$
with the sections (\ref{sect}).
The elements $u_j$ are the points of the Jacobian  $Jac(\Si_\tau)$.
They play the role of the coordinates, and thereby, $C_\tau\sim Jac(\Si_\tau)$.

This gauge fixing is invariant with respect to the constant diagonal subgroup $D_0$.
It acts on the spin variables $S\in\clO$.
This action is Hamiltonian. The moment equation of this action is $\di(\clO)=0$.
This condition dictates the form of $S^0=J$. The gauge fixing allows one to kill the
degrees of freedom related to the spin variables, because $\dim(\clO)=2(N-1)$ and
$\dim(D_0)=N-1$. Thus, the symplectic quotient is a point $(\dim(\clO//D_0)=0)$.
\begin{rem}
One can choose an arbitrary orbit $\clO$. In this case we come to the symplectic
quotient $\clO//D_0$. It has dimension $\dim(\clO)-2(N-1)$.
\end{rem}

 Now consider solutions the moment equation (\ref{gal}) with the prescribed
monodromies and prove that $\Phi$ becomes
the Lax matrix
$\Phi\to L^{CM}=V+X$ (\ref{lcm}). Since $\bA_{0}=0\,$, $\,V$ does not contribute in (\ref{gal}) and its
elements are free parameters. We identify them with momenta of the particles
$V=\di(v_1,\ldots,v_N)$.
Due to the term with the delta-function in  (\ref{gal}) the off-diagonal part
should has a simple pole with the residue $\nu J$ and the prescribed
monodromies.
It follows from (\ref{phi}) and (\ref{resi}) that $X_{jk}$ satisfies these conditions.
They uniquely fix its matrix elements.

The reduced space is described by the variables $\bfv$ and $\bfu$.
The symplectic form on the reduced space
$$
\int_{\Si_{1,1}}\lan L^{CM},\bA_{0}\ran=\sum Dv_j\wedge Du_j
$$
leads to the brackets  $\{v_j,u_k\}=\de_{jk}$.

From the general construction the integrals of motion come from the expansion
of\\
$\tr (L^{CM})^j(\bfv,\bfu,z)\,,$. They are double periodic
meromorphic functions
with poles at $z=0$. It is finite-dimensional space generated by a basis of
derivative of the Weierschtrass functions. They are elements of the basis
$\varsigma^{jk}$ in (\ref{int2}).
\beq{exp}
\f1{j}\tr(L^{CM})^j(\bfv,\bfu,z)=I^{CM}_{0,j}+I^{CM}_{2,j}\wp(z)+
\dots+I^{CM}_{j,j}\wp^{(j)}(z)\,.
\eq
There are $\frac{N(N+1)}{2}-1$ integrals. Due to a special choice of the orbit
only $N-1$ integrals are independent.
In particular,
$$
\oh\tr(L^{CM})^2(\bfv,\bfu,z) =-H^{CM}+\nu^2\wp(z)\,.
$$

For generic orbits (see Remark 4.1) the Hamiltonian take the form
$$
H^{CM}=\oh|\bfv|^2+\sum_{j<k}S_{jk}S_{kj}\wp(u_j-u_k)\,.
$$
It is the ECMS with spin \cite{GH}.
Note, that $I_{j,j}$ are the Casimir functions defining a generic orbit $\clO$.
Therefore we have $\frac{N(N+1)}{2}-1-(N-1)=\frac{N(N-1)}{2}$ commuting integrals
of motion. The number of independent commuting integrals is always equal to $\oh\dim(\clO)$.

%%%%%%%%%%%%%%%%%%%%%%%%%%%%%%%%%%%%%%%%%%%%%%%%%%%%%%%%%%%%%%%%%%%%%%%%%%%%%%%%%%%

\subsection{Elliptic Top (ET) on $\GLN$}

{\bf 1.Description of system}

The elliptic top is an example of Euler-Arnold top related to the group $\GLN$.
Its phase space is a coadjoint orbit of $\GLN$. The Hamiltonian is a quadratic form
on the coalgebra $\gg^*=\gln^*$. The ET is an integrable Euler-Arnold top.
Before define the Hamiltonian introduce a special basis in the Lie algebra
 $\gln$. Define the finite set
$$
\mZ^{(2)}_N=(\mZ/N\mZ\oplus\mZ/N\mZ)\,,~~
\ti{\mZ}^{(2)}_N=(\mZ/N\mZ\oplus\mZ/N\mZ)\setminus(0,0)\,
$$
and let $\bfe_N(x)=\exp \frac{2\pi i}N x$. Then a basis is generated by
$N^2-1$ matrices
$$
 T_{\al}=
\frac{N}{2\pi i}\bfe_N(\frac{\al_1\al_2}{2})Q^{\al_1}\La^{\al_2}\,,~
\al=(\al_1,\al_2)\in\ti{\mZ}^{(2)}_N\,,
$$
where
\beq{qq}
Q=\di(1,\bfe_N(1),\ldots,\bfe_N(N-1))\,,
\eq
\beq{la}
\La=\sum_{j=1,N,~(mod~N)}E_{j,j+1}\,.
\eq
The commutation relations in this basis have a simple form
$$
[T_{\al},T_{\be}]=
\frac{N}{\pi}\sin\frac{\pi}{N}(\al\times \be)T_{\al+\be}\,.
$$
Let $\bfS=\sum_{\al\in{\mZ}^{(2)}_N\setminus(0,0)}S_\al T_\al\in\gg^*$.
The Poisson brackets for the linear functions $S_\al$ come from the Lie brackets
$$
\{S_\al,S_\be\}=
\frac{N}{\pi}\sin\frac{\pi}{N}(\al\times\be)S_{\al+\be}\,.
$$

The phase space ${\cal R}^{ET}$ of the ET
is a coadjoint orbit
$$
{\cal R}^{ET}\sim{\cal O}=\{\bfS\in\gg^*\,|\,\bfS=
g\bfS_0 g^{-1}\,,~g\in\GLN\}\,.
$$
A particular orbit passes through $\bfS_0=\nu J$, as for the spinless ECMS.

The Euler-Arnold Hamiltonian is defined by the quadratic form
$$
H^{ET}=-\oh\tr( \bfS\cdot\bfJ(\bfS))\,,
$$
where $\bfJ$ is diagonal in the basis $T_\al$
$$
\bfJ(\bfS)~:~S_\al\to \wp_\al S_\al\,,~~
\wp_\al=\wp\left(\frac{\al_1+\al_2\tau}N\right)\,,~
\al\in\ti{\mZ}^{(2)}_N\,.
$$
The equations of motion corresponding to this Hamiltonian take the form
$$
\p_t\bfS=\{H^{ET},\bfS\}=[\bfJ(\bfS),\bfS]\,,
$$
$$
\p_t S_\al=\frac{N}{\pi}
\sum_{\ga\in\ti{\mZ}^{(2)}_N}S_\ga S_{\al-\ga}\wp_\ga
\sin\frac{\pi}{N}(\al\times \ga)\,.
$$

\bigskip
%%%%%%%%%%%%%%%%%%%%%%%%%%%%%%%%%%%%%%%%%%%%%%%%%%%%%%%%%%%%%%%%%%%%%%%%%%%%%%%%%%%%%%%%%%%

{\bf 2.Field theory and the Higgs bundles}

The curve $\Si_{1,1}$ is the same as
for the Calogero-Moser system.
 Consider a vector bundle $E$ of a rank $N$ and degree one
  over $\Si_{1,1}$. It is described by its sections
$s=(s_1(z,\bz),\ldots,s_N(z,\bz))$   with monodromies
\beq{qp}
s^T(z+1,\bz+1)=Q^{-1}s^T(z,\bz)\,,~~s^T(z+\tau,\bz+\bar{\tau})=\ti{\La}^{-1}s^T(z,\bz)\,,
\eq
where $Q$ is (\ref{qq}),
$\ti{\La}=\bfe_N^{-(z+\frac{\tau}{2})}\La$,
and $\La$ is (\ref{la}).
Since $\det Q=\pm 1$ and $\det\ti{\La}=\pm \bfe_1^{-(z+\frac{\tau}{2})}$
the determinants of the transition matrices have the same quasi-periods
as the Jacobi theta-functions. The  theta-functions have a simple pole
in $\Si_{1,1}$. Thereby, the vector bundle $E_N$ has degree one.

  The Higgs bundle has the same field content as the ECMS
$$
\mathcal{R}=\{\bA,\Phi,\bfS\}\,,~~\bA,\Phi\in\gln\,,~~
\bfS\in\mathcal{O}\,.
$$
The orbit
$$
\clO=\{\bfS=g^{-1}\bfS^0g\,, ~~g\in \GLN\}
$$
is located  at the marked point $z=0$.

It follows from (\ref{qp}) that the fields $\Phi$, $\bA$ have the
monodromies
$$
\bA(z+1)=Q\bA(z)Q^{-1}\,,~~
\Phi(z+1)=Q\Phi(z)Q^{-1}\,,
$$
$$
\bA(z+\tau)=\La \bA(z)\La^{-1}\,,~~
\Phi(z+\tau)=\La\Phi(z)\La^{-1}\,.
$$

The group of the automorphisms $\clG_\mC=\{f\}$ of  $E$ should have the same monodromies
$$
f(z+1)=Qf(z)Q^{-1}\,,~~f(z+\tau)=\La f(z)\La^{-1}\,.
$$

Due to the monodromy conditions
the generic field $\bA$
is gauge equivalent to the trivial $f^{-1}\bA f+f^{-1}\bp f=0$.
Therefore
\beq{pg}
\bA=-\bp f[\bA]f^{-1}[\bA]\,.
\eq
 It allows us to choose $\bA=0$ as an appropriate gauge. It means that there are
 no moduli of holomorphic vector bundles. More precisely, the holomorphic moduli
 are related only to the quasi-parabolic structure of $E$ related to the spin
 variables $\bfS$.
The monodromies of the gauge matrices prevent to
have nontrivial residual gauge symmetries. Let $f[\bA](z,\bz)$ be
a solution of (\ref{pg}).
Consider the transformation of $\Phi$ by solutions of (\ref{pg})
\beq{lr}
L^{ET}[\bA,g](z,\bz)=f[\bA](z,\bz)\Phi(z,\bz)f^{-1}[\bA](z,\bz)\,.
\eq
The moment constraints (\ref{HE}) takes the form
$$
\bp L^{ET}=\de(z,\bz)\bfS\,.
$$
The solution takes the form
$$
\fbox{$ L^{ET}=\sum_{\al\in{\mZ}^{(2)}_N\setminus(0,0)}S_\al
\varphi_\al(z) T_\al\,,$}
$$
where
$\varphi_\al(z)=\bfe_N(\al_2z)\phi(\frac{\al_1+\al_2\tau}{N},z)$.
The Lax matrix was found in Ref.\,\cite{RSTS}) using another approach.
It is the Lax matrix of the vertex spinchain.
The Lax matrix is meromorphic on $\Si_{1,1}$ with a simple pole
with $Res\,L^{ET}|_{z=0}=\bfS$.
The monodromies of $\varphi_\al(z)$ are read off from (\ref{phi})
$$
\varphi_\al(z+1)=\bfe_N(\al_2)\varphi_\al(z)\,,~
~\varphi_\al(z+\tau)=\bfe_N(-\al_1)\varphi_\al(z)\,.
$$
Then $L^{ET}$ has the prescribed monodromies.
The reduced phase space $\clR^{ET}$ is the coadjoint orbit:
$$
{\cal R}^{ET}=\{{\cal O}=\bfS=g\bfS_0g^{-1}\}\,,
$$
$\bfS=\sum_{\al\in{\mZ}^{(2)}_N\setminus(0,0)}S_\al T_\al\in\gg^*$.
The symplectic form on ${\cal R}^{ET}$ is the Kirillov-Kostant form (\ref{s2}).

For a particular choice of the orbit passing through $J$ (ref{J}) its dimension coincide with the
dimension of the phase of the spinless ECMS
$$
\dim{\cal R}^{ET}=\dim{\cal R}^{CMS}=2N-2\,.
$$
It is not occasional and we prove below that ${\cal R}^{CM}$ is symplectomorphic
to ${\cal R}^{ET}$.

Since the traces $\tr (L^{ET})^j$ are double periodic and have poles at $z=0$
the integrals of motion come from the expansion (see (\ref{exp}))
$$
\tr(L^{ET}(z))^k=I_{0,k}+I_{2,k}\wp(z)+\ldots+I_{k,k}\wp^{(k-2)}(z)\,.
$$
In particular,
$$
\tr (L^{ET})^2=H^{ET}+C^2\wp(z)\,.
$$
The coefficients $I_{s,k}$ are in involution
$$
\{I_{s,k},I_{m,j}\}=0\,.
$$
In particular, all functions $I_{s,k}$ Poisson commute with the
Hamiltonian $H^{ET}$. Therefore,
they play the role of conservation laws of elliptic rotator
hierarchy on $\GLN$.
We have a tower of $\frac{N(N+1)}{2}$ independent integrals of motion
$$
\begin{array}{ccccc}
I_{0,2}&I_{2,2}&      &  & \\
I_{0,3}&I_{2,3}&I_{3,3}&  & \\
\ldots&\ldots&\ldots&\ldots&\\
I_{0,n}&I_{2,N}&\dots&\dots&I_{N,N}\\
\end{array}
$$
There is no integrals $I_{1,k}$ because there is no double periodic
meromorphic functions with one simple pole. The integrals
$I_{k,k},~k=0,2,3\ldots,N$ are the Casimir functions corresponding to the
coadjoint orbit
$$
{\mathcal R}^{ET}=\{\bfS\in\gln,~~\bfS=g^{-1}\bfS^{(0)}g\}\,.
$$
The conservation laws $I_{s,k}$ generate commuting flows on ${\cal R}^{rot}$
$$
\p_{s,k}\bfS=\{I_{s,k},\bfS\}_1\,, ~~(\p_{s,k}:=\p_{t_{s,k}})\,.
$$

%%%%%%%%%%%%%%%%%%%%%%%%%%%%%%%%%%%%%%%%%%%%%%%%%%%%%%%%%%%%%%%%%%%%%%%%%%%%%%%%%%%%%%%%%%%%

%%%%%%%%%%%%%%%%%%%%%%%%%%%%%%%%%%%%%%%%%%%%%%%%%%%%%%%%%%%%%%%%%%%%%
%%%%%%%%%%%%%%%%%%%%%%%%%%%%%%%%%%%%%%%%%%%%%%%%%%%%%%%%%%%%%%%%%%%%%%%%%%%%%%%%%%%%%%%%%%%%

\subsection{Symplectic Hecke correspondence}

Let $E$ and $\ti{E}$ be two bundles over $\Si$
of the same rank. Assume that there is a map
$\Xi^+~ \colon E\to \tilde{E}$ (more precisely
a map of the space of sections $\G(E)\to\G(\tilde{E})$)   such that
it is an isomorphism on the complement to
$z_0$ and it has one-dimensional cokernel at $x\in\Sigma$~:
$$
0\to E\stackrel{\Xi^+}{\rightarrow}
\tilde{E}\to \Bbb C|_{z_0}\to 0\,.
$$
The map $\Xi^+$ is called {\it upper modification} of the bundle $E$ at the point $z_0$.
Let $w=z-z_0$ be a local coordinate in a neighborhood of $z_0$.
We represent locally $E$ as a sum of line bundles $E=\oplus_{j=1}^N\clL_j$
with holomorphic sections
\beq{sec}
s=(s_1,s_2,\ldots,s_N)\,.
\eq
 After the modification we come to the bundle
$\ti{E}=\oplus_{j=1}^N\clL_j\otimes\clO(z_0)$.
The sections of $\ti{E}$ are represented  locally as $\ti{s}=(g_1(w)s_1,\ldots, w^{-1}g_N(w)s_N)$, where $g_j(0)\neq 0$.
In this basis
the upper modification at the point $z_0$ is represented by the matrix
$$
\Xi^+=\mat{{\rm Id}_{N-1}}{0}{0}{w}\,.
$$
It is a modification of order 1, since it increase the degree of $E$
\beq{degr}
{\rm deg}\,(\ti{E})={\rm deg}\,(E)+{\rm deg}\,(\clO(z_0)={\rm deg}\,(E)+1\,.
\eq
On the complement to the point $z_0$ consider the map
$$
E\stackrel{~\Xi^-}\leftarrow\ti{E}\,,
$$
such that $\Xi^-\Xi^+=$Id. It defines {\it the lower modification}
at the point $z_0$. The upper modification $\Xi^+$ is represented by the
vector $(0,\ldots,1)$ and $\Xi^-$ by  $(0,\ldots,-1)$.

For the Higgs bundles the modification acts as
$$
(E,\Phi)\stackrel{\Xi}{\rightarrow}
(\tilde{E},\ti{\Phi})
$$
\beq{him}
\Xi\Phi=\ti{\Phi}\Xi\,,~~\Xi\ti{\bA}=\bp\Xi+\bA\Xi\,.
\eq
The Higgs fields $\Phi$ and $\ti{\Phi}$ should be holomorphic with prescribed simple poles at the marked points. The holomorphity of the Higgs field put restrictions on its form.
Consider the upper modification $\Xi^+\sim(0,\ldots ,1)$ and assume that $\Phi$ in the defined above basis takes the form
$$
\Phi=\mat{a}{b}{c}{d}\,,
$$
where $a$ is a matrix of order $N-1$.
Then
$$
\Xi\mat{a}{b}{c}{d}=\mat{a}{bw^{-1}}{cw}{d}\Xi\,.
$$
We see that a generic Higgs field acquire a first order pole after the modification.
To escape it we assume that there exists  an eigen-vector
$\Phi\xi=\la\xi$ such that it belongs to the $Ker\Phi$. Let $\xi=(0,0,\ldots,1)$ and
$$
\Phi=\mat{a}{0}{c}{d}\,.
$$
Then the Higgs field $\ti{\Phi}$ does not have a pole
$$
\ti{\Phi}=\mat{a}{0}{cw}{d}\,.
$$
In other words the matrix elements $(\Phi)_{jN}$ should have first order null.

In this way the upper modification is lifted from $E$ to the Higgs bundle. After
the reduction we come to the map (see (\ref{mhb}))
$$
T^*\clM(N,g,n,d)\to T^*\clM(N,g,n,d+1)\,.
$$
We call it the upper {\it Symplectic Hecke Correspondence} (SHC).

 Generically the modified bundle $\ti{E}$ is represented locally
as a sum of line bundles\\
 $\ti{E}=\oplus_{j=1}^N(\clL_j\otimes\clO(z_0)^m_j\,$
$\,(m_j\in\mZ)$ with holomorphic
sections
\beq{ms1}
\ti{s}=(\ti s_1,\ldots,\ti s_N)=(w^{-m_1}g_1s_1,w^{-m_2}g_2s_2,\ldots,w^{-m_N}g_Ns_N)\,.
\eq
 It has degree
$$
{\rm deg}\,(\ti{E})={\rm deg}\,(E)+\sum_{j=1}^Nm_j\,.
$$
This modification is represented by the vector $(m_1,\ldots,m_n)$.

Remind that
the Higgs field is an endomorphism of $E\,$  $\,s\to\Phi s\,$ and near $z_0$ it acts as
$$
\Phi\cdot s_j=(\Phi)_j^ks_k\,.
$$
Similarly the modified Higgs field acts on sections of the modified bundle $\ti{E}\,$ $\ti s\to\ti\Phi\ti{s}$.
Then it follows from (\ref{ms1}) that
$$
\ti\Phi\cdot \ti s_j=\ti \Phi_j^k\ti s_k\,,~~~ \ti \Phi_j^k=w^{m_k-m_j}g_k(w)g_j^{-1}(w)\Phi_j^k\,.
$$
Since  $\,\ti\Phi$ is holomorphic and $g_j(0)\neq 0\,$, $\Phi_j^k(z-z_0)^{m_k-m_j}$ must
be regular at $z=z_0$. If we order\\
$m_1\geq m_2\geq\ldots\geq m_N$ then the number
of parameters of the endomorphisms is $\sum_{j<k}(m_j-m_k)$.
In general case
$$
T^*\clM(N,g,n,d)\to T^*\clM(N,g,n,d+\sum_{j=1}^Nm_j)\,.
$$
 If $\sum_{j=1}^Nm_j=0$ the SHC does not change the topological type of the bundle.
Therefore, such SHC defines a  B$\rm \ddot{a}$cklund transformation of integrable
hierarchy.

%%%%%%%%%%%%%%%%%%%%%%%%%%%%%%%%%%%%%%%%%%%%%%%%%%%%%%%%%%%%%%%%%%%%%
%%%%%%%%%%%%%%%%%%%%%%%%%%%%%%%%%%%%%%%%%%%%%%%%%%%%%%%%%%%%%%%%%%%%%%%%%%%%%%%%%%%%%%%%%%%%

\subsection{Symplectic Hecke correspondence ${\mathcal R}^{CM}\to {\mathcal R}^{ET}$. \cite{LOZ1}}

We work directly with the Lax matrices
$$
L^{ET}\times\Xi=\Xi\times L^{CM} \,.
$$
 The modification matrix should intertwine
the multipliers  corresponding to the fundamental cycles
  \beq{mo1}
\Xi(z+1,\tau)= Q\times \Xi(z,\tau)\,,
\eq
\beq{mo2}
\Xi(z+\tau,\tau)=\tilde\Lambda(z,\tau)\times
\Xi(z,\tau)
\times{\rm diag}
({\bf e}(u_j))\,.
\eq

Consider the modification at $z=0$. The Lax
matrix of the CMS has the first order pole
$$
L^{CM}\sim\f1{z}\nu J\,.
$$
Its residue has an eigen-vector $\xi^t=(1,\ldots,1)$ with the eigen-value $N-1$. The matrix $\Xi$ satisfying (\ref{mo1})
and (\ref{mo2}) that annihilates the vector $\xi$ has the form
$$
\Xi(z)=\tilde\Xi(z)\times{\rm diag}\left((-1)^l
\prod_{j<k;j,k\ne l}
\theta(u_k-u_j,\tau)\right)
$$
$$
\tilde\Xi_{ij}(z, u_1,\ldots,u_N;\tau) =
\theta{\left[\begin{array}{c}
\frac{i}N-\frac12\\
\frac{N}2
\end{array}
\right]}(z-Nu_j, N\tau )\,.
$$
Here $\theta{\left[\begin{array}{c}
\frac{i}N-\frac12\\
\frac{N}2
\end{array}
\right]}(z-Nu_j, N\tau )$ is the theta-function with a characteristic.
The determinant of $\Xi$ can be calculated explicitly
$$
\det\left[\frac{\tilde\Xi_{ij}(z, u_1,\ldots,u_N;\tau)}
{i\eta(\tau)}\right]=
\frac{\theta(z)}{i\eta(\tau)}\prod\limits_{1\leq k<l\leq N}
\frac{\theta(u_l-u_k)}{i\eta(\tau)}\,,
$$
where $\eta(\tau)=q^{\frac{1}{24}}\prod_{n>0}(1-q^n)$
is the Dedekind function. It has a simple pole at $z=0$ and therefore $\Xi$
is degenerate.

We use the modification to write down the interrelations between the coordinates and momenta
of the Calogero-Moser particles and the orbit variables of the Elliptic Top in the
SL$(2,\mC)$ case
$$
S_1=-v\frac{\theta_{10}(0)}{\theta'(0)}
\frac{\theta_{10}(2u)}{\theta(2u)}
-
\nu\frac{\theta_{10}^2(0)}{\theta_{00}(0)\theta_{01}(0)}
\frac{\theta_{00}(2u)\theta_{01}(2u)}{\theta^2(2u)}\,,
$$
\beq{s3}
S_2=-v\frac{\theta_{00}(0)}{i\theta'(0)}
\frac{\theta_{00}(2u)}{\theta(2u)}
-
\nu\frac{\theta_{00}^2(0)}{i\theta_{10}(0)\theta_{01}(0)}
\frac{\theta_{10}(2u)\theta_{01}(2u)}{\theta^2(2u)}\,,
\eq
$$
S_3=-v\frac{\theta_{01}(0)}{\theta'(0)}
\frac{\theta_{01}(2u)}{\theta(2u)}
-
\nu\frac{\theta_{01}^2(0)}{\theta_{00}(0)\theta_{10}(0)}
\frac{\theta_{00}(2u)\theta_{10}(2u)}{\theta^2(2u)}\,.
$$
Here
$\theta_{1,0}=\sum_{n\in\mZ}q^{\oh n^2}\exp\pi(2n-1)z\,$,
$\,\theta_{0,0}=\sum_{n\in\mZ}q^{\oh(n-\oh)^2}\exp 2\pi nz$,\\
$\theta_{0,1}=\sum_{n\in\mZ}(-1)^nq^{\oh(n-\oh)^2}\exp 2\pi nz\,.$
These relations describe the Darboux coordinates $(v,u)\in\mC^2$ the coadjoint
${\rm SL}(2,\mC)$-orbit $\sum S_\al^2=\nu^2$

It turns out that this modification is equivalent to the twist of $R$-matrices.
Namely, it describes the passage from the  dynamical $R$ matrix of the IRF models to
the vertex $R$-matrix \cite{B,JMO}. We don't discuss this aspect of SHC here.

%%%%%%%%%%%%%%%%%%%%%%%%%%%%%%%%%%%%%%%%%%%%%%%%%%%%%%%%%%%%%%%%%%%%%%%%%%%
%%%%%%%%%%%%%%%%%%%%%%%%%%%%%%%%%%%%%%%%%%%%%%%%%%%%%%%%%%%%%%%%%%%%%%%%%%%
\part{Lecture 3}
\section{4d theories}
\setcounter{equation}{0}

\subsection{Self-dual YM equations and Hitchin equations}

\subsubsection{2-d self-dual equations}
Consider a rank $N$ complex vector bundle $E$ over  $\mR^4$ with coordinates
$\bfx=(x_0,x_1,x_2,x_3)$.
Assume that the space of sections is equipped with a nondegenerate Hermitian metric
$h\,$,\\
 $(h^+=h)$. It satisfies the following condition
 $dh(x,y)=h(\nabla x,y)+h(x,\nabla y)$, where $\nabla$ is a connection on $E$.
 If $dh(x,y)=0$ for vectors in fibers
 $y\in V,$ $\,x\in\bar{V}^t$, then there exist connections $\nabla_j=\p_{x_j}+A_j$
such that
$$
\bfA^+=-h^{-1}dh-h^{-1}\bfA h\,,~~~(\bfA=\sum_{j=0}^3A_jdx_j)\,.
$$
In this situation the transition functions are reduced to the unitary group
$\SUN\subset\GLN$.

Let $F(\bfA)\in \Om^{(2)}(\mR^4,\sun) $ be the curvature
$F_{ij}=[\nabla_i,\nabla_j]$ or $F(\bfA)=d\bfA+\bfA^2$.
Here \\
$\sun=\{x\,|\,x^+=-h^{-1}xh\}$

The self-duality equation
$$
F=\star F\,,
$$
where $\star$ is the Hodge operator in $\mR^4$
takes the form
\beq{sd}
\left\{
\begin{array}{c}
F_{01}=F_{23}\\
F_{02}=F_{31}\\
F_{03}=F_{12}
\end{array}
\right.
\eq

Assume that $A_j$ depend only on $(x_1,x_2)$. It means that
the fields are invariant under the shifts in directions $x_0,x_3$.
Then $(A_0,A_3)$ become adjoint-valued
scalar fields which we denote  as $(\phi_1,\phi_2)$.
They are called  the Higgs fields. In fact, they will be associated
below with the Higgs field $\Phi$.
 In this way we come
to the self-dual equations on the plane $\mR^2=(x_1,x_2)$
\beq{sd1}
F_{12}=[\phi_1,\phi_2]\,,
\eq
\beq{sd2}
[\nabla_1,\phi_1]= [\phi_2,\nabla_2]\,,
\eq
\beq{sd3}
[\nabla_1,\phi_2]=[\nabla_2,\phi_1]\,.
\eq

Introduce  complex coordinates
$z=x_1+ix_2\,,$ $\,\bz=x_1-ix_2$ and let
$\,d'=\nabla_z$, $\,d''=\nabla_{\bz}$. Consider
the fields, taking values in the Lie algebra $\sln$
$$
\left\{
\begin{array}{c}
\Phi_z=\oh(\phi_1-i\phi_2)dz\in\Om^{(1,0)}(\mR^2,\ad\, E)\,,\\
 \Phi_{\bz}=\oh(\phi_1+i\phi_2)d\bz\in\Om^{(0,1)}(\mR^2,\ad\, E)\,.
 \end{array}
 \right.
$$
They are not independent since the Hermitian conjugation acts as
\beq{uni}
\Phi_{\bz}^+=-h^{-1}\Phi_zh\,.
\eq
Similarly,
$$
\left\{
\begin{array}{c}
A_z=\oh(A_1-iA_2)\\
 A_{\bz}=\oh(A_1+iA_2)\,,
 \end{array}
 \right.
$$
\beq{uni1}
A_{\bz}^+=-h^{-1}dh-h^{-1}A_zh\,.
\eq

In terms of  fields
\beq{W}
\clW=(A\,,A_{\bz}\,,\Phi_z\,,\Phi_{\bz})
\eq
 (\ref{sd1}) -- (\ref{sd3}) can be rewritten in the coordinate invariant way:
\beq{he}
\left\{
\begin{array}{ll}
1.\,& F+[\Phi_z,\Phi_{\bz}]=0\,,\\
2.\,& d''\Phi_z=0\,,\\
3.\,& d'\Phi_{\bz}=0\,,
\end{array}
\right.
\eq
where $[\Phi_z,\Phi_{\bz}]=\Phi_z\Phi_{\bz}+\Phi_{\bz}\Phi_z\,$.
Due to (\ref{uni}) and  (\ref{uni1}) the third equation is not independent.
Thus, we have two equations with the left side of type $(1,1)$ for two complex valued fields
$(\Phi_z,A_{\bz})$ and the hermitian matrix $h$.

The equations (\ref{he}) are conformal invariant and thereby can be defined
on a complex curve $\Si_g$.
In this case
$$
\Phi_z\in\Om^{(1,0)}(\Si_g,\sun)\,,~~\Phi_{\bz}\in\Om^{(0,1)}(\Si_g,\sun)\,,
$$
$$
d''\,:\,\Om^{(j,k)}(\Si_g,\sun)\to\Om^{(j,k+1)}(\Si_g,\sun)\,.
$$
The self-duality equations (\ref{he}) on  $\Si_g$
are called \emph{the Hitchin equations}.

Consider  the gauge group  action on solutions of (\ref{he})
\beq{gt3}
\clG=\{f\in\Om^{0}(\Si_g,\SUN)\}\,,
\eq
\beq{gt1}
\Phi_z\to f^{-1}\Phi_z f\,,~~\Phi_{\bz}\to f^{-1}\Phi_{\bz} f\,,
\eq
\beq{gt2}
d''\to f^{-1}d''f\,.
\eq
If $(A\,,A_{\bz}\,,\Phi_z\,,\Phi_{\bz})$ are solutions of  (\ref{he}), then
the transformed fields are also solutions.
If $f$ takes values in $\GLN$  then it again transforms solutions to solutions.
As above we denote this gauge group as $\clG_{\mC}$.

Define the moduli space of solutions of (\ref{he}) as a quotient under the gauge
group action
\beq{ms}
\clM_H(\Si_g)={\rm solutions~of~(\ref{he})}/\clG\,.
\eq

Now look on the second equation in (\ref{he}). It is the moment constraint equation for the Higgs bundles in the absence of marked points (\ref{HE}).
The gauge group  $\clG_{\mC}$ transforms solutions of  (\ref{he}) to solutions but breaks (\ref{uni}), (\ref{uni1}). Now restrict ourself with the second equation in (\ref{he}).
Dividing the space of its solution on the gauge group $\clG_\mC$
we come to the moduli space of the Higgs bundles $T^*\clM(N,g,0,d)$ (\ref{mhb}).
There exists a dense subset of moduli space of stable Higgs bundles
$(T^*\clM(N,g,0,d))^{stable}\subset T^*\clM(N,g,0,d)$.
 The moduli space of  stable Higgs bundles parameterize the smooth part of $\clM_H(\Si_g)$ (\ref{ms}) \cite{H2}.

Consider a Higgs bundle with a data $(\Phi,\bA)$ satisfying eq. 2 in (\ref{he})
and reconstruct from it solutions $(A_z,\Phi_z,A_{\bz},\Phi_{\bz})$  of (\ref{he}).
Define them as
$$
\Phi_z=\Phi\,,~~~\Phi_{\bz}=-h^{-1}\Phi^+h\,,
$$
$$
A_{\bz}=\bA\,,~~~A_z=-h^{-1}\bp h-h^{-1}\bA^+h\,.
$$
Then $(\Phi_{\bz},A_z)$ satisfy eq. 3.(\ref{he}). The equation 1.(\ref{he}) takes the form
$$
\bp(h^{-1}\bp h+h^{-1}\bA^+h)-
\p\bA+[\bA,(h^{-1}\bp h+h^{-1}\bA^+h)]-[\Phi,h^{-1}\Phi^+h]=0\,.
$$
For almost all $(\Phi,\bA)$ there exists a solution $h$ of this equation
(see appendix of Donaldson in \cite{H2}).
In this way we pass from the holomorphic data to solutions of system (\ref{he}).

Summarizing, to define $\clM_H(\Si_g)$ one can acts in two ways:\\
1.  Divide the space of solutions of (\ref{he}) on the $\SUN$-valued gauge group
$\clG$.\\
2. Consider the moduli space of stable Higgs bundles.

%%%%%%%%%%%%%%%%%%%%%%%%%%%%%%%%%%%%%%%%%%%%%%%%%%%%%%%%%%%%%%%%%%%%%

\subsubsection{Hyper-Kahler reduction}

In this section we explain how to derive the moduli space $\clM_H(\Si_g)$ (\ref{ms})
 via an
analog of the symplectic reduction. It is a so-called Hyper-Kahler reduction
\cite {HKLR}.
We prove that infinite-dimensional space $\clW$ (\ref{W})
is a Hyper-Kahler manifold, and
 $\clM_H$ is its Hyper-Kahler quotient,
where (\ref{he}) play the role of the moment equations.

To define a  Hyper-Kahler manifold we need a three complex structures and
a metric satisfying certain axioms.
Define a flat metric on $\clW$ depending on the
complex structure on $\Si$
\beq{4.10}
ds^2=-\f1{4\pi}\int_{\Si}Tr(\de A_z\otimes\de A_{\bz}+\de A_{\bz}\otimes\de A_z+
\de\Phi_z\otimes\de \Phi_{\bz}+\de\Phi_{\bz}\otimes\de\Phi_z)\,.
\eq
Introduce three complex structures $I\,,J\,,K$ on $\clW$. The
corresponding operators act on the tangent bundle $T\clW$,
such that they obey the imaginary quaternion relations $I^2=J^2=K^2=-Id\,$,
$\,IJ=K\,,\ldots$. The complex structures are integrable because $\clW$ is flat.
Introduce a basis of one-forms in $T^*\clW$
$$
V=(\de A_{\bz}\,,\de\Phi_z\,,\de A_z\,,\de\Phi_{\bz})\,.
$$
Then the action of the conjugated operators on $T^*\clW$ in this basis takes the form
$$
I^T=\left(
  \begin{array}{cccc}
    i & 0 & 0 & 0 \\
    0 & i & 0 & 0 \\
    0 & 0 & -i & 0 \\
    0 & 0 & 0 & -i \\
  \end{array}
\right)\,,~~
J^T=\left(
                \begin{array}{cccc}
                  0 & 0 & 0 & -1 \\
                  0 & 0 & 1 & 0  \\
                  0 & -1 & 0 & 0 \\
                  1 & 0 & 0 & 0 \\
                \end{array}
              \right)\,,~~
K^T=\left(
    \begin{array}{cccc}
      0 & 0 & 0 & -i \\
      0 & 0 & i & 0\\
       0 & i & 0 & 0 \\
        -i & 0 & 0 & 0 \\
    \end{array}
  \right)\,.
  $$
Linear functions on $\clW$ are holomorphic with respect to a complex structure, if they are transformed under the action of the corresponding operator with
eigen-value $+i$. Thus $A_z$, $\Phi_z$ are holomorphic in the complex structure $I$,
$A_{\bz}+i\Phi_{\bz}$, $A_z+i\Phi_z$ are holomorphic in the complex structure $J$, and
$A_{\bz}-\Phi_{\bz}$, $A_z+\Phi_z$ are holomorphic in the complex structure $K$.

To be hyper-Kahler on $\clW$ the metric $ds^2$ should be of type $(1,1)$ in each complex structure. It means that
$ds^2\sim (I^T\otimes I^T)ds^2=(J^T\otimes J^T)ds^2=(K^T\otimes K^T)ds^2$.
In this way we have described a flat hyper-Kahler metric on $\clW$.
A linear combination of the complex structures produces a family of
complex structures, parameterized by $\mC\mP^1$.

We define three symplectic structures associated with the complex structures
on $\clW$ as $\om_I=(I^T\otimes Id)ds^2$, $\,\om_J=(J^T\otimes Id)ds^2$,
$\,\om_K=(K^T\otimes Id)ds^2$.
$$
\om_I=-\frac{i}{2\pi}\int_{\Si_g}\tr(D A_{\bz}\wedge D A_z-D\Phi_z\wedge D\Phi_{\bz})\,,
$$
\beq{ss}
\om_J=\f1{2\pi}\int_{\Si_g}\tr(D\Phi_{\bz}\wedge
D A_z+D\Phi_z\wedge\de A_{\bz})\,,
\eq
$$
\om_K=\frac{i}{2\pi}\int_{\Si_g}\tr(D\Phi_{\bz}\wedge D A_z-D\Phi_z\wedge D A_{\bz})\,.
$$
These forms are closed and of type $(1,1)$ with respect to the corresponding complex structures.

Now consider the gauge transformations (\ref{gt3}) of the fields (\ref{gt1}),
(\ref{gt2}). Since the gauge transform takes values in $\SUN$,
 the forms (\ref{ss}) are gauge invariant. Therefore we can proceed  as
in the case  of the standard symplectic reduction (\ref{mom}).
But now we obtain three generating momentum Hamiltonians with respect to the three symplectic forms
$$
F_I=-\frac{i}{2\pi}\int_{\Si_g}\tr(\ep(F_{z,\bz}-[\Phi_z,\Phi_{\bz}]))\,, ~~(\ep\in Lie(\clG))\,,
$$
$$
F_J=-\f1{2\pi}\int_{\Si_g}\tr(\ep(d'\Phi_{\bz}+d''\Phi_z))\,,
$$
$$
F_K=-\frac{i}{2\pi}\int_{\Si_g}\tr(\ep(d'\Phi_{\bz}-d''\Phi_z))\,.
$$
and the three moment maps $\clW\to Lie^*(\clG)$
$$
\mu_I=F-[\Phi_z,\Phi_{\bz}]\,,~~\mu_J=d'\Phi_{\bz}+d''\Phi_z\,,
~~\mu_K=i(d'\Phi_{\bz}-d''\Phi_z)\,.
$$
The zero-valued moments coincide with the Hitchin systems.
The hyper-Kahler quotient $\clW///\clG$ is defined as
$$
\clW///\clG=\mu_I^{-1}(0)\cap\mu_J^{-1}(0)\cap\mu_K^{-1}(0)/\clG\,.
$$
To come to the system (\ref{he}) consider the linear combination
\beq{nui}
\nu_I=\mu_J+i\mu_K=d''\Phi_{\bz}\,.
\eq
This moment map is derived from the symplectic form
$$
\Om_I=\om_J+i\om_K=\f1{\pi}\int_\Si\tr(D\Phi_z\wedge D A_{\bz})\,.
$$
It is a $(2,0)$-form in the complex structure $I$. Thus we have the holomorphic moment
map $\nu_I$ in  the complex structure $I$. Vanishing of the holomorphic moment map
$\nu_I$ and the real moment map $\mu_I$ is equivalent to the Hitchin equations.
Dividing their solutions on the gauge group $\clG$ we come to
 the moduli space $\clM_H(\Si_g)$ (\ref{ms}).

Now consider an analog of (\ref{nui}) corresponding to the complex structure $J$
$$
\nu_J=\mu_K+i\mu_I=\clF_{z,\bz}\,,~~~~\clF_{z,bz}=\clF(\clA_z,\clA_{\bz})\,,
$$
$\clA_z=A_z+i\Phi_z$, $\,\clA_{\bz}=A_{\bz}+i\Phi_{\bz}$.
This moment map comes from the symplectic form
$$
\Om_J=\f1{2\pi}\int_{\Si_g}\tr(D \clA\wedge D \clA)\,.
$$
It is $(2,0)$ form in the complex structure $J$.
Putting $\nu_J=0$ we come to flatness condition of the bundle $E$.
Dividing the set of solutions $\clF_{z,\bz}=0$ on the $\GLN$ valued gauge
transformations $\clG_\mC$ we come to the space
\beq{cly}
\clY=(\clF_{z,bz}=0)/\clG_\mC
\eq
of homomorphisms $\pi_1(\Si_g)\to\GLN$ defined up to conjugations.
According with
\cite{Co} and Donaldson (the appendix in Ref.\,\cite{H2}) generic flat bundles
parameterize $\clM_H(\Si_g)$ (\ref{ms}) in the complex structure $J$.
This space is a phase space of non-autonomous Hamiltonian systems leading
to monodromy preserving equations (see Section {\bf 6.3}).
Thus, the space  $\clM_H(\Si_g)$ describes  phase spaces of integrable
systems $\clR^{red}$ (\ref{mhb}) in the complex structure $I$ and  phase spaces of monodromy preserving equations $\clY$ (\ref{cly}) in the complex structure $J$.

%%%%%%%%%%%%%%%%%%%%%%%%%%%%%%%%%%%%%%%%%%%%%%%%%%%%%%%%%%%%%%%%%%%%%%%%%%%%%%%%%%%%%%%%%%%%%%%%%%%%%%
%%%%%%%%%%%%%%%%%%%%%%%%%%%%%%%%%%%%%%%%%%%%%%%%%%%%%%%%%%%%%%%%%%%%%%%%%%%%%%%%%%%%%%%%%%%%%%%%%%%%%%

\subsection{$\mathcal{N}=4$ SUSY Yang-Mills in four dimension
and Hitchin equations}

Here we consider a twisted version of $\mathcal{N}=4$ super Yang-Mills theory in
four dimension. This theory was analyzed in detail in \cite{KW,GW,W} to develop a
field-theoretical approach to
the Geometric Langlands Program. The quantum Hitchin systems is a one side of this construction
and we use here only a minor part of  \cite{KW}.
The twisted theory is a topological theory that contains a generalization of the
Hitchin equations (\ref{he}) as a condition of the BRST invariance. Our goal is to
 describe the Hecke transformations in terms of the theory.
In section 4 we have defined the Hecke transformations as an instant singular gauge transformation. The four-dimensional theory allows to consider gauge transformations varying  along a space coordinate $x_3$. They become singular at some point, say $x_3=0$, where a singular t'Hooft operator
is located. It gives  a natural description of the symplectic Hecke correspondence in
terms of a monopole configuration in the twisted theory.

%%%%%%%%%%%%%%%%%%%%%%%%%%%%%%%%%%%%%%%%%%%%%%%%%%%%%%%%%%%%%%%%%%%%%%%%%%%%%%%%%%%%%%%%%%%%

\subsubsection{Twisting of $\mathcal{N}=4$ SUSY $\SUN$ Yang-Mills theory}

$\mathcal{N}=4$ SUSY $\SUN$ Yang-Mills action in four dimension can be derived  from
the $\mathcal{N}=1$ SUSY $\SUN$ Yang-Mills action in ten dimensions by
the dimensional reduction. We need only the bosonic part of the reduced theory.

The bosonic fields of the 4d Yang-Mills theory are four-dimensional gauge potential
$$
\bfA=(A_0\,,A_1\,,A_2\,,A_3)\,,
$$
and six scalar fields coming from six extra dimensions
$$
\bfphi=(\phi_0\,,\phi_1\,,\phi_2\,,\phi_3,\phi_4,\phi_5)\,.
$$
The bosonic part of the action has the form
$$
I=\f1{e^2}\int d^4x\tr\left(\oh\sum_{\mu,\nu=0}^3F_{\mu\nu}F^{\mu\nu}
+\sum_{\mu=0}^3\sum_{i=1}^6D_\mu\phi_iD^\mu\phi_i+\oh
\sum_{i,j=1}^6[\phi_i\phi_j]^2\right)\,.
$$
The symmetry of the action is  $Spin(4)\times Spin(6)$ (or $Spin(1,3)\times Spin(6)$
in the Lorentz signature). The sixteen generators of the 4d  supersymmetry
are transformed under $Spin(1,3)\times Spin(6)$ $\sim$
$SL(2)\times SL(2)\times Spin(6)$ as $(2,1,\bar{4})\oplus(1,2,4)$:
$$
\{\bar{Q}_{AX}\}\oplus\{Q_{\dot{A}}^Y\}\,,
~~(A=1,2;\,X=1,\dots,4)\,,~(\dot{A}=1,2;\,Y=1,\dots,4)\,.
$$
They satisfy the super-symmetry algebra
\beq{q}
\{\bar{Q}_{AX},Q_{\dot{A}}^Y\}=\de_X^Y\sum_{\mu=0}^3\Gamma^\mu_{A\dot{A}}P_\mu
\eq
$$
\{Q,Q\}=0\,,~~\{\bar{Q},\bar{Q}\}=0\,.
$$
The action of $Q$ on a field $X$ takes the form
$$
\de X=[Q,X\}\,.
$$

Let $\ka$ be a map $Spin(4)\,\to\, Spin(6)$ and set
$$
Spin'(4)=(Id\times \ka)
$$
Define $\ka$ in such a way that the action of $Spin'(4)$ on the  chiral spinor $\clS^+$
has an invariant vector. Let $Q$ be the  corresponding supersymmetry. It follows
from (\ref{q}) that it obeys $Q^2=0$. The twisted theory is defined by the physical observables
from the cohomology groups $H^\bullet(Q)$.
The twisted four scalar fields $\phi=(\phi_0,\dots,\phi_3)$ are reinterpreted as
adjoint-valued one-forms on $\mR^4$, while untwisted
$\sigma,\bar{\sigma}=\phi_4\pm\imath\phi_5$ remain adjoint-valued scalars.

In fact there is a family of topological theories parameterized by $t\in\mC\mP^1$.
The bosonic fields to be invariant under $Q$ should satisfy the equations
\beq{ft}
\begin{array}{ll}
  1)&\,(F-\phi\wedge\phi+tD\phi)^+=0\,, \\
   \\
  2)& \,(F-\phi\wedge\phi-t^{-1}D\phi)^-=0\,,\\
   \\
   3)&\,\star D\star\phi=0\,,
\end{array}
\eq
where $^\pm$ denote the self-dual and the anti-self-dual parts for four-dimensional
two-forms,\\
$D=d+[\bfA,~]$ and $\star$ is the Hodge operator in four dimension.
We are interesting in solutions of this system up to gauge
transformations.

This theory defined on flat $\mR^4$ can be extended on any four-manifold $M$
in such a way that it preserves the $Q$-symmetry and contributions of metric come
 only from $Q$-exact terms.
 The bosonic part of the theory is described by connections $\bfA=(A_0,A_1,A_2,A_3)$
 in a bundle $E$ over $M$ in a presence of the adjoint-valued one-forms
 $\bfphi=(\phi_0,\phi_1,\phi_2,\phi_3)$  satisfying (\ref{ft}).

The important for integrable systems case is $M=\mR^2\times\Si_g$,
where\\
 $\mR^2=(time=x_0)\times \{x_3=y\}$
and $\Si_g$ will play the role of the basic spectral curve. $\mR^2$ is not involved
in the twisting and the  fields $(\phi_0,\phi_3)$ remain  scalars, while $\phi_1,\phi_2$ become
one-forms on $\Si_g$. It turns out that after the reduction the system (\ref{ft}) becomes
equivalent to the Hitchin equations (\ref{he}).

%%%%%%%%%%%%%%%%%%%%%%%%%%%%%%%%%%%%%%%%%%%%%%%%%%%%%%%%%%%%%%%%%%%%%%%%%%%%%%%%%%%%%%%%%%%

\subsubsection{Hecke correspondence and monopoles}

The system (\ref{ft}) for $t=1$ can be replaced by
\beq{1}
F-\phi\wedge\phi+\star D\phi=0\,,
\eq
\beq{2}
\star D\star\phi=0\,.
\eq
Assume that the fields are time independent and consider the system on
the three-dimensional manifold
$W=I(x_3)\times\Si_g$, where $-\infty\leq x_3\leq \infty$.
In terms of the tree-dimensional
fields $\tilde{A}$ and $\tilde{\Phi}$
$\,(\bfA=(A_0,\ti{A})\,,~\phi=(\Phi_0 dx_0,\tilde{\Phi}))$,
 the equations take the form
\beq{ft1}
\begin{array}{l}
  \tilde{F}-\tilde{\Phi}\wedge\tilde{\Phi}=\star(D\Phi_0-[A_0,\tilde{\Phi}])\,, \\
  \\
  \star D\tilde{\Phi}=[\Phi_0,\tilde{\Phi}]+DA_0\,, \\
  \\
  \star D\star\tilde{\Phi}+[A_0,\Phi_0]=0\,.
\end{array}
\eq
Here the Hodge operator $\star$ is taken in the three-dimensional sense.
Replace the coordinates $\vec{x}=(x_1,x_2,x_3)\,$ on  $x_3\to y$  and $(x_2,x_3)\to(z,\bz)$,
where  $(z,\bz)$ are local coordinates  on $\Si_g$.
Let $g(z,\bz)|dz|^2$ be a metric on $\Si_g$. Then the metric  on $W$ is   $ds^2=g|dz|^2+dy^2$. Then the Hodge operator takes the form
$$
\star dy=\oh i gdz\wedge d\bz\,,~
\star dz= -i dz\wedge dy\,,~
\star d\bz= i d\bz\wedge dy\,.
$$
It can be found that $\phi_y=0$ and $A_0=0$ are solutions of the system.
 Taken the gauge $A_y=0$ we come to the equations
\beq{ft2}
\fbox{$
\begin{array}{ll}
 1.  &F(A_z,A_{\bz})-[\Phi_z,\Phi_{\bz}]=\oh i g\p_{y}\Phi_0\,. \\
 2.  & D_{A_{\bz}}\Phi_z=0\,, \\
3. & \p_yA_{\bz}=-i D_{A_{\bz}}\Phi_0\,,   \\
4. & \p_y\Phi_{z}=-i [\Phi_{z}\Phi_0]\,,\\
\end{array}
$}
\eq
where as before $\Phi_z^+=-\Phi_{\bz}$.
It follows from 3. and 4. that the scalar field $\Phi_0$ plays the role of a
gauge transformation. For $\Phi_0=0$  the system  (\ref{ft2}) becomes
essentially two-dimensional and  coincides with
the Hitchin equations  (\ref{he}).

Let $\Si_g$ be an elliptic curve $(g=1)$. This case is important to
application to integrable systems. The nonlinear system (\ref{ft2}) can be rewritten
as a compatibility condition for the linear system depending on the spectral parameter
$\la\in\mC$
$$
\left\{
\begin{array}{l}
(\p_z+\la^{-1}a\p_y+A_z+i\la^2\Phi_z-i\la^{-1}a\Phi_0)\psi=0\,,\\
(\p_{\bz}+\la a\p_y+A_{\bz}+i\la^{-2}\Phi_{\bz}+i\la a\Phi_0)\psi=0\,
\end{array}
\right.
$$
Here $a^2=-\frac{i}{\bar{\tau}-\tau}$. This linear system allows to apply the
methods of the Inverse Scattering Problem
or the Whitham approximation to find solutions of  (\ref{ft2}).

Now assume that $\Phi_0(z,\bz,y)$ is  non-zero. It preserve $\Phi_z=0$.
In this case the first equation in (\ref{ft2}) is the Bogomolny equation
\beq{3}
F(A_z,A_{\bz})=\star \p_y\Phi_0\,.
\eq
Consider a monopole solution of this equation. Let $\ti{W}=(W\setminus \vec{x}^0=(y=0,z=z_0))$.
The Bianchi identity $ DF=0$ in the space $\ti{W}$ implies that
$\Phi_0$ is the Green function for the operator $\star D\star D$
$$
\star D\star D\Phi_0=\de(\vec{x}-\vec{x}^0)\,.
$$

Consider first the abelian case $G=$U$(1)$. Then $F(A_z,A_{\bz})$ is a curvature of
a line bundle $\clL$.
Locally near $\vec{x}_0=(y=0,z=z_0,\bz=\bz_0)\,$  $\,\Phi_0$ has a singularity
\beq{4}
\Phi_0\sim\frac{im}{2|\vec{x}-\vec{x}^0|}\,,
\eq
and since 1. in (\ref{ft2}) takes the form
$$
F(A_z,A_{\bz})=\oh  g\p_y\Phi_0\,.
$$
$$
F(A_z,A_{\bz})\sim\oh  mg(z,\bz)\frac{y}{|\vec{x}-\vec{x}^0|^{\frac{3}{2}}}\,.
$$
Consider a small sphere $S^2$ enclosing $\vec{x}^0$. Due to (\ref{3}) and (\ref{4})
$$
\int_{S^2}F=m\,.
$$
This solution describes the Dirac monopole of charge $m$ corresponding to a line bundle over
$S^2$ of degree $m$.

Let $\Si_g^\pm=\Si_g\times (\pm \infty)$ and $\clL^\pm$ be the line bundles
over $\Si_g^\pm$. The two-dimensional cycle $C$ describing the boundary
$C=\p((W=I\times\Si_g)\setminus\vec{x}_0)$ is
$\Si_g^+-\Si_g^--S^2$. Taking the integral over $C$ we find that
$$
\int_C F=0\,.
$$
In other words, for the Chern classes of the bundles $c(\clL)=deg(\clL)$ we have
$$
deg (\clL^+)=deg(\clL^-)+m\,,
$$
or $\clL^+=\clL^\otimes\clO(z_0)^m$. Here $\clO(z_0)^m$ is a line bundle whose
holomorphic sections are holomorphic functions away from $z_0$ with a possible single
pole of degree $m$ at $z_0$. The line bundles over $\Si_g$ are topologically equivalent
for $y<0$ or $y>0$. The gauge transformation $\Phi_0$ is smooth away from $\vec{x}_0$.
The singularity change the degree of the bundle.

\bigskip

\includegraphics{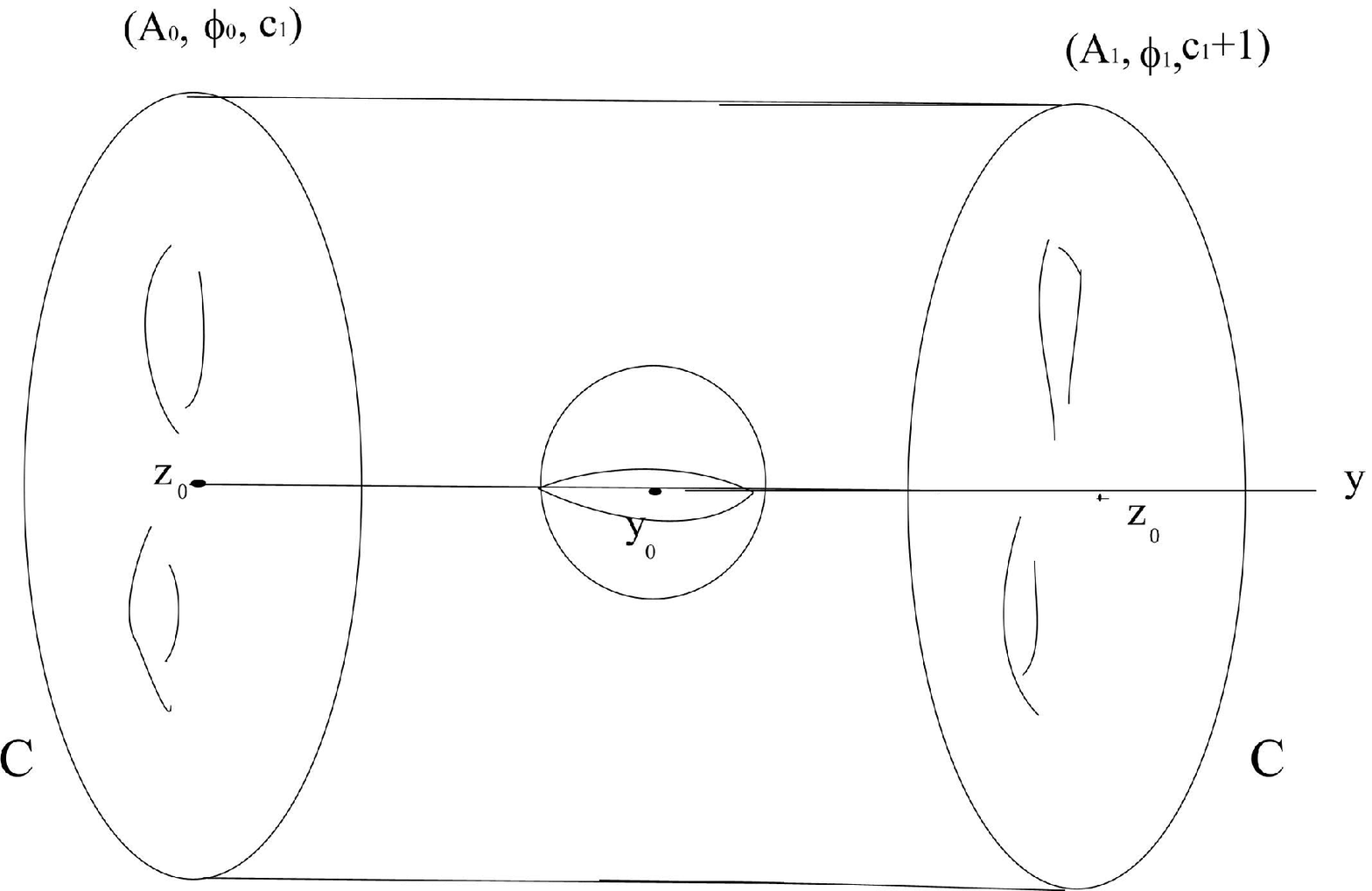}
The monopole increase the Chern class: $c_1\to c_1+m$

\bigskip

In the four-dimensional abelian theory we have the  Dirac monopole
singular along the time-like line $L=(x_0,\vec{x}^0)$. It corresponds to including
\emph{the t'Hooft operator} in the theory saying that the connections have
the monopole singularity along the line $L$.

A generic vector bundles $E$ near $\vec{x}_0$ splits
$E_y\sim \clL_1\oplus\clL_2\oplus\ldots\oplus\clL_N$.
Consider the gauge transformation
\beq{gtr}
\Phi_0\sim\frac{i}{2|\vec{x}-\vec{x}_0|}\di(m_1,\ldots,m_N)\,.
\eq
It causes the transformation $\clL_j\to\clL_j\otimes\clO(z_0)^{m_j}$.
The degree of the bundles $E$ changes after crossing $y=0$ by $\sum m_j$,
as it was described for bundles over $\Si$ in Section {\bf 4.4}.

To be more precise we specify the boundary conditions of solutions on the ends $y=-\infty$
and $y=+\infty$. Since $\Phi_0\to 0$ for $y\to\pm\infty$ the system (\ref{ft2}) coincides with
the Hitchin system (\ref{he}). If $\clM_H(N,g,n,m^\pm)$ is the moduli space of solutions on
the boundaries $y=\pm\infty$ the gauge transformation with the monopole singularity stands that
$m^+=m^-+\sum m_j$. It defines the SHC between two integrable systems related to
$\clM_H(N,g,n,m^\pm)$. In particular, we have described it at the point $y=0$ for
$\clM_H(N,1,n,0)$ and $\clM_H(N,1,n,1)$.

\section{Conclusion}

Here we shortly discus some related issues have not included in the  lectures.

\bigskip

\begin{description}
  \item[1] Solutions of the Hitchin equations (\ref{he}) corresponding to
quasi-parabolic Higgs bundles were analyzed in Ref.\,\cite{GW}.
In the three-dimensional gauge theory considered in Section 4.3 we have the Wilson lines located at the marked points.
  In the four-dimensional Yang-Mills theory they corresponds to singular operators
along two-dimensional surfaces. Locally on a punctured disc around a
marked point the Hitchin system (\ref{he}) assumes the form of the Nahm equations \cite{N}.
It was proved in Ref.\,\cite{Kro} that the space of its solutions
after dividing on a special gauge group is symlectomorphic to a coadjoint orbit of $\SLN$.
A hyper-Kahler structure on the space of solutions induces a hyper-Kahler structure on the orbits. It establishes the interrelations between the Hitchin equations  and the Higgs bundles
with the marked points (the quasi-parabolic Higgs bundles).

  \item[2] There exists a generalization of this approach to  Higgs bundles
  of infinite rank.
  In other words, the structure group $G=\GLN$ or $\SLN$ of the bundles
   is replaced by an infinite-rank group. One way is to consider
 % Let $G=\hat{L}(\SLN)$ be
  the central extended loop group $S^1\to G$. Then the Higgs
  field depends on additional variable $x\in S^1$ and
   instead of
  the Lax equation we come to the Zakharov-Shabat equation
  $$
  \p_jL-\p_xM_j+[M_j,L]=0\,.
  $$
  This equation describes an infinite-dimensional integrable hierarchy like the
  KdV hierarchy. The two-dimensional version of the ECMS was constructed in \cite{Kr1,LOZ1}. In particular, the SHC
  establishes an equivalence of the two-particles $(N=2)$
  elliptic Calogero-Moser field theory with the Landau-Lifshitz equation
  \cite{Sk,Bo}.
  The latter system is the two-dimensional version of the SL$(2,\mC)$ elliptic top.
  The relations (\ref{s3}) are working in the two-dimensional case.

  Another way is to consider GL$(\infty)$ bundles. In Ref.\,\cite{O} the  ECMS for
  infinite number of particles $N\to\infty$ was analyzed. The elliptic top on the group
  of the non-commutative torus was considered in Ref.\,\cite{KLO}. It is a subgroup
  of GL$(\infty)$. This construction describes an integrable modification
  of the hydrodynamics of the ideal fluid on a non-commutative two-dimensional torus.

  \item[3]
  Consider dynamical  systems, where the role of times is played by
  parameters of complex structures of curves $\Si_{g,n}$. In this case we come to
   monodromy preserving equations, like the Schlesinger system or the
  Painlev{\'e} equations. They can be constructed in the similar
  fashion as the integrable Hitchin systems \cite{LO}.
  To this purpose the one should replace the Higgs bundles by the flat bundles and
  afterwards use the same symplectic reduction (see (\ref{cly})). In this situation the Lax equations
  takes the form
  $$
  \p_jL-\p_zM_j+[M_j,L]=0\,.
  $$
An analysis of this system is more complicated in compare with the standard Lax equations
due to the presence of derivative with respect to the spectral parameter.
Note that $M_j$ corresponds only to the quadratic Hamiltonians, since they responsible for
the deformations of complex structures.
Concrete examples of this construction was given in \cite{LO,Ta1,CLOZ}. Interrelations
with Higgs bundles were analyzed  in \cite{LO,Ta}.
It is remarkable that the Symplectic Hecke correspondence is working in this case.
It establishes an equivalence of the Painlev{\'e} VI equation and a non-autonomous
Zhukovski-Volterra gyrostat \cite{LOZ2}.

\item[4]
A modification of the Higgs bundles allows one to construct relativistic integrable systems
\cite{Ru}.
The role of Higgs field is played by a group element $g=\exp (c K^{-1}\Phi)$
where $K$ is a canonical class on $\Si$ and $c$ is the relativistic parameter.
 This construction is working only for curves of genus $g\leq 1$.
This approach was realized in Ref.\,\cite{AFM} to derive the elliptic Rujesenaars
system and in Ref.\,\cite{BDOZ,CLOZ} to derive the elliptic classical r-matrix of
Belavin-Drinfeld \cite{BD} and a quadratic Poisson algebra of the
Sklyanin-Feigin-Odesski type \cite{Skl,FO}.

Including the relativistic systems allows to define a duality in integrable systems
\cite{FGNR,GR} (see \cite{AP} for recent developments). This type of dualities has
a natural description for the corresponding quantum integrable systems in terms
of Hecke algebras \cite{Ch}. It is called there the Fourier transform and
takes the form of
$S$-duality. Another form of duality in the classical Hitchin system considered
in \cite{HT,DP,Hi3}. It is related to Langlands duality and similar to $T$-duality of fibers in the Hitchin fibration.

\item[5]
There exists an useful description of the moduli space of holomorphic vector bundles
closely related to the modification described in Section 4.4. It is so-called Tyurin
parametrization \cite{Tyu}. This construction was applied to describe Higgs bundles and
integrable systems related to curve of arbitrary genus in Ref.\,\cite{Kr1,ER1,Ta1}.
Using this approach  classical r-matrices with a spectral parameter living on curves
of arbitrary genus was constructed in Ref.\,\cite{Dol}.

\end{description}

\section{Bibliography}

\small{

}

\end{document}